\documentclass[prd,twocolumn,english,preprintnumbers,amsmath,amssymb,nofootinbib,superscriptaddress,a4]{revtex4-1}

\usepackage[latin1]{inputenc}
\usepackage{graphicx}
\usepackage{color}

\usepackage{color}

\usepackage{bm}
\usepackage{bbm}
\usepackage{amsmath,amsfonts,latexsym}

\usepackage{dsfont}

\newcommand{\Eqref}[1]{Eq.~\eqref{#1}}

\newcommand{\xcm}{x_{\text{CM}}}
\newcommand{\nL}{n_{\text{L}}}

\usepackage{babel}
\makeatother

\begin{document}

\title{Vacuum polarization tensor in inhomogeneous magnetic fields}
\author{Holger Gies and Lars Roessler}
\affiliation{Theoretisch-Physikalisches Institut, Friedrich-Schiller-Universit\"at Jena,  Max-Wien-Platz 1,
             D-07743 Jena, Germany, \\
             \& Helmholtz Institut Jena,  Helmholtzweg 4,
             D-07743 Jena, Germany, 
} 

\begin{abstract}
  We develop worldline numerical methods, which combine string-inspired with
  Monte-Carlo techniques, for the computation of the vacuum polarization
  tensor in inhomogeneous background fields for scalar QED. The algorithm
  satisfies the Ward identity exactly and operates on the level of
  renormalized quantities. We use the algorithm to study for the first time
  light propagation in a spatially varying magnetic field. Whereas a local
  derivative expansion applies to the limit of small variations compared to
  the Compton wavelength, the case of a strongly varying field can be
  approximated by a derivative expansion for the averaged field.  For rapidly
  varying fields, the vacuum-magnetic refractive indices can exhibit a
  non-monotonic dependence on the local field strength. This behavior can
  provide a natural limit on the self-focussing property of the quantum
  vacuum.
\end{abstract}

\maketitle

\section{Introduction}

The study of vacuum polarization has been one of the cornerstones of the
development of quantum field theory. It gives access to charge and field
strength renormalization and provides for an understanding of
fluctuation-induced modifications of Coulomb's law in terms of the Uehling
potential which contributes to the Lamb shift. 

Vacuum polarization in strong magnetic fields has lead to the prediction of
nonlinear optical properties of the quantum vacuum
\cite{Toll:1952,Baier,Adler:1971wn,Tsai:1975iz,Gies:1999vb}, for a review, see
\cite{Dittrich:2000zu}. More specifically, the magnetized quantum vacuum is
birefringent for low-energy photons and exhibits dichroism above the threshold
for pair production \cite{Tsai:1974fa,Daugherty:1984tr}. Past and present
experiments, such as BFRT \cite{Cameron:1993mr}, PVLAS
\cite{Zavattini:2007ee}, BMV \cite{Robilliard:2007bq}, Q\&A
\cite{Chen:2006cd}, OSQAR \cite{Pugnat:2006ba}, have been working on the
discovery of these elementary properties of the quantum vacuum by means of
strong macroscopic magnetic fields. In addition to being a fundamental test of
QED, these experiments can provide for the strong laboratory bounds on the
existence of hypothetical particles such as minicharged
or axion-like particles 
\cite{Maiani:1986md,Gies:2006ca,Ahlers:2006iz,Ahlers:2007rd}. 

As first suggested in \cite{Heinzl:2006xc}, also high-intensity lasers in
combination with latest methods in X-ray polarimetry \cite{Marx:2011ab} may
contribute to the search and discovery of these nonlinear properties of the
quantum vacuum. Various laser-induced quantum-vacuum phenomena have been
investigated recently
\cite{DiPiazza:2005jc,Schutzhold:2008pz,Tommasini:2009nh,King:10ab,Blaschke:2011af,Harvey:2009ry},
for reviews see \cite{Marklund:2008gj,Dunne:2008kc,Heinzl:2008an}. They 
also provide for a powerful probe for new elementary particles complementary
to accelerators \cite{Gies:2008wv,Homma:2010jc,Dobrich:2010hi} or supplement
searches for a noncommutative structure of spacetime \cite{Heinzl:2009zd}.

Theoretical analyses of the vacuum polarization tensor so far
dealt with the idealized limit of constant, homogeneous fields
\cite{Batalin:1971au,Ritus:1972ky,Cover:1974ij,Tsai:1974fa,Tsai:1975iz,%
Urrutia:1977xb,Artimovich:1990qb,Gies:1999vb,Schubert:2000yt,%
Dittrich:2000wz,Konar:2001mp,VillalbaChavez:2010bp} which is
appropriate as long as the scale of 
field variation is much larger than the Compton wavelength in QED. Whereas this is
well satisfied for experiments with large dipole fields searching for effects
from fluctuations of standard model particles, this assumption can be violated
for secondary fields (e.g., from higher harmonics) from strong lasers or for
fluctuations involving hypothetical very light particles. A proper
interpretation of future experiments will therefore require the knowledge of
the vacuum polarization tensor in general inhomogeneous fields. 

Recent years have witnessed a variety of advances for strong-field
calculations in inhomogeneous fields. So far, new techniques have concentrated
mainly on the effective action or effective Lagrangian in strong-fields as a
primary quantity of interest. In addition to exact solutions
\cite{Nikishov:1970br,Cangemi:1995ee}, semiclassical \cite{Kim:2000un},
instanton techniques, and quantum kinetic equations have been developed and
applied to pair production in inhomogeneous fields, i.e., the imaginary part
of the action, as reviewed in \cite{Dunne:2008kc}. As a general-purpose
numerical method, the combination of the worldline formalism
\cite{Halpern:1977he,Schubert:2001he} with Monte-Carlo path integration
techniques has proved successful in many instances
\cite{Gies:2001zp,Langfeld:2002vy,Gies:2005bz,Gies:2003cv}. As one advantage,
worldline numerics is capable of providing {\em local} information about the
effects of fluctuations, such as energy or action densities, local production
rates etc. Local quantities are particularly indicative for the non-local
features of quantum-field theory.

The present work is devoted to generalizing these worldline methods to the
vacuum polarization tensor as the lowest nontrivial correlation function of
QED. As the effective action is the generating functional for 1PI
correlation functions, the generalization at first sight seems
straightforward. However, the relation is provided by a functional
differentiation which is difficult to implement reliably in a numerical
method. In addition to being a powerful numerical method, the success of
numerical worldline techniques also relies on the fact that the formulation is
very close to analytical calculations. In fact, closed-form worldline expressions for
correlation functions to arbitrarily high order (master formulas) can be
derived within perturbation theory \cite{Schubert:2001he}. 

In this work, we demonstrate that the corresponding formula for the vacuum
polarization tensor in an external field can be used to develop a worldline
numerical algorithm that (i) satisfies the Ward identities at any level of
discretization, and (ii) operates on the level of renormalized quantities so
that only finite quantities are subject to numerical evaluation. For numerical
simplicity, we perform all computations within scalar QED -- the
generalization to spinor QED within the worldline approach is
straightforward \cite{Langfeld:2002vy,Dunne:2009zz}. 

As a concrete example, we concentrate on the evaluation of the refractive
indices of the magnetized quantum vacuum deduced from the vacuum polarization
tensor and their dependence on spatial variations.  This is not only a
crucial observable in birefringence experiments, but can describe a
particular sensitivity of optical observables to nonlocal features of the
quantum vacuum. For instance, as the vacuum refractive index in the
homogeneous field approximation increases with the field strength, the quantum
vacuum has a self-focussing property \cite{Kharzeev:2006wg}: photons are drawn into
regions of higher field strength, in turn amplifying the field
strength even more. Our results provide for first indications that the
refractive index can depend non-monotonically on the field strength in regions
of large spatial variations, hence providing for a natural mechanism to limit
this self-focussing property of the quantum vacuum. 

This article is organized as follows: in Sect.~\ref{sec:II}, we summarize the
approach to the polarization tensor on the worldline. We pay special attention
to spacetime inhomogeneities and develop a numerical algorithm which satisfies
the Ward identity at any level of discretization. Sect.~\ref{sec:III} is
devoted to benchmark tests in the form of comparisons with analytically known
vacuum and constant-field cases. In Sect.~\ref{sec:IV}, we present new results
for the polarization tensor in an inhomogeneous magnetic field. 


\section{Vacuum polarization tensor on the worldline}
\label{sec:II}

Let us start with the worldline representation of the one-loop effective
action of scalar QED in $D$ Euclidean spacetime dimensions \cite{Schubert:2001he},
\begin{align*}
	\Gamma[\mathcal{A}] =
        \int\limits_{0}^{\infty}\frac{dT}{T}\frac{e^{-m^2T}}{(4\pi
          T)^{D/2}}\int\limits_{x(0) =
          x(T)}\mathcal{D}x~e^{-\int_{0}^{T}d\tau~(\frac{\dot{x}^2}{4}+ie\dot{x}\mathcal{A})} ,
\end{align*}
where $m$ denotes the (scalar) electron mass. A transition to
Minkowski-valued quantities will be discussed below. Here, the path integral
is normalized to unity for vanishing field $\mathcal{A}=0$. Ultraviolet
divergencies can be regularized with the help of the propertime $T$ integral,
e.g., by replacing the lower integration limit $T=0$ by
$T_{\text{min}}=1/\Lambda^2$ (propertime cutoff), or by dimensional or zeta
function regularization. The corresponding effective action for spinor QED
looks very similar, additionally containing a spin-field coupling
\cite{Schubert:2001he}. This effective action is the generating functional for
all 1PI correlation functions, which can be deduced from $\Gamma$ by
functional differentiation with respect to the gauge field
$\mathcal{A}_\mu$. Alternatively, we can expand the gauge field in terms of a
background field $A^\mu$ and a sum over plane waves, $\mathcal{A}_{j}^{\mu}(x)
= A^\mu+ \sum\limits_{j = 0}^{\infty}\epsilon_{j}^{\mu}e^{ik_{j}x}$. The
second order is relevant for the vacuum polarization tensor,
\begin{widetext}
\begin{align*}
	\Gamma &=
        (-ie)^2\int\limits_{0}^{\infty}\frac{dT}{T}\frac{e^{-m^2T}}{(4\pi
          T)^{D/2}}\int\limits_{x(0) =
          x(T)}\mathcal{D}x~e^{-\int_{0}^{T}d\tau~\frac{\dot{x}^2}{4}}
        \left\{\int\limits_{0}^{T}d\tau_{1}
        \int\limits_{0}^{T}d\tau_{2}~\dot{x}_{1}\epsilon_{1}e^{ik_{1}x_{1}}
        \dot{x}_{2}\epsilon_{2}e^{ik_{2}x_{2}}
        e^{-ie\oint dx A} \right\}
        + \mathcal{O}(\epsilon^3)\\ 
	&= \epsilon_{1\mu}\Gamma^{(2)\mu\nu}[k_{1}, k_{2};A]\epsilon_{2\nu}
        + \mathcal{O}(\epsilon^3).
\end{align*}
%
%
The plane wave basis also implements the transition to momentum space. At
first sight, this appears less efficient, as the worldline integrals live
in position space; an evaluation of the 2-point correlator in position space
thus seems to be much more straightforward. However, it turns out that the
position space representation naturally involves two path integrals (one for
each internal propagator), whereas the momentum space formulation boils down
to one path integral and thus is is numerically less expensive.  The desired
vacuum polarization tensor can be extracted as (part of) the coefficient of the
polarization vectors $\epsilon_j^\mu$,
%
\begin{align*}
	\Gamma^{(2)\mu\nu}[k_{1}, k_{2};A] &=
        (-ie)^2\int\limits_{0}^{\infty}\frac{dT}{T}\frac{e^{-m^2T}}{(4\pi
          T)^{D/2}}\int\limits_{x(0) =
          x(T)}\mathcal{D}x~e^{-\int_{0}^{T}d\tau~\frac{\dot{x}^2}{4}}\left\{\int\limits_{0}^{T}d\tau_{1}\int\limits_{0}^
          {T}d\tau_
          {2}~\dot{x}_{1}^{\mu}e^{ik_{1}x_{1}}\dot{x}_{2}^{\nu}e^{ik_{2}x_{2}}
          e^{-ie\oint dx A}  \right\}.
\end{align*}
%
%
The worldline integral can be decomposed into a path integral over worldlines
with a common center of mass $\xcm$ and the spacetime integration over this
center of mass\footnote{Different prescriptions for such a decomposition can
  equally well be used \cite{Schubert:2001he,Dunne:2009zz}.}
%
\begin{align*}
\int\mathcal{D}x~e^{-\int_{0}^{T}d\tau~\frac{\dot{x}^2}{4}}&\left\{\int\limits_{0}^{T}d\tau_{1}\int\limits_{0}^
          {T}d\tau_
          {2}~\dot{x}_{1}^{\mu}e^{ik_{1}x_{1}}\dot{x}_{2}^{\nu}e^{ik_{2}x_{2}}
          e^{-ie\oint dx A}  \right\}\\
	&= \int\limits_{x(0) =
          x(T),\text{CM}}\mathcal{D}x~e^{-\int_{0}^{T}d\tau~\frac{\dot{x}^2}{4}}\left\{\int\limits_{0}^{T}d\tau_{1}\int\limits_{0}^
          {T}d\tau_	{2}~\dot{x}_{1}
          {\mu}e^{ik_{1}x_{1}}\dot{x}_{2}^{\nu}e^{ik_{2}x_{2}}\int
          d^{D}\xcm~e^{i(k_{1}+k_{2})\xcm} e^{-ie\oint dx A}\right\}.\\
\end{align*}
%
In the case of a homogeneous background or in the vacuum case, there is no
additional $\xcm$ dependence apart from the plane waves, such that the $\xcm$
integration can immediately be performed, yielding a $\delta$ function that
implements momentum conservation, see below. In the general case, we have to
be more careful and introduce a {\em local} vacuum polarization tensor $\pi^{\mu\nu}$ that
depends on $\xcm$,
\begin{equation}
	\Gamma^{(2)\mu\nu}[k_{1}, k_{2};A] =\int
        d^{D}\xcm~ e^{i(k_{1}+k_{2})\xcm}
        \pi^{\mu\nu}[k_1,k_2,\xcm;A],
\label{eq:defpi}
\end{equation} 
where the local polarization tensor is
%
%
%
\begin{equation}
\pi^{\mu\nu}[k_1,k_2,\xcm;A]
	= (-ie)^2\int\limits_{0}^{\infty}\frac{dT}{T}\frac{e^{-m^2T}}{(4\pi
          T)^{D/2}}
        \int\limits_{x(0) = x(T),
          \text{CM}}\!\!\!\!\!\!\! 
        \mathcal{D}x~e^{-\int_{0}^{T}d\tau~\frac{\dot{x}^2}{4}}
        \left\{\int\limits_{0}^{T}d\tau_{1}\int	\limits_{0}^{T}d\tau_	{2}
          ~\dot{x}_{1}^{\mu}e^{ik_1x_{1}}\dot{x}_{2}^{\nu}e^{-ik_2x_{2}}e^{-ie\oint dx
            A} \right\}.\label{eq:defpi1}
\end{equation}

\end{widetext}
In the homogeneous or vacuum case, \Eqref{eq:defpi} boils down to
\begin{equation}
	\Gamma^{(2)\mu\nu}[k_{1}, k_{2};A] =(2\pi)^D  \delta^{(D)}(k_1+k_2) \Pi^{\mu\nu}[k;A],
\label{eq:defPi}
\end{equation}
where $\Pi^{\mu\nu}[k;A]$ is the standard vacuum polarization tensor in a
homogeneous field, and $k=k_1=-k_2$. Momentum conservation is automatically
implemented, and local and global descriptions are identical.

U(1) gauge symmetry imposes a constraint on the polarization tensor in the
form of the Ward identity. In its local form, the polarization tensor
has to satisfy
\begin{equation}
k_{1,\mu} \pi^{\mu\nu}[k_1,k_2,\xcm;A] = \pi^{\mu\nu}[k_1,k_2,\xcm;A]
k_{2,\nu} =0. 
\label{eq:WI}
\end{equation}
In the worldline representation \eqref{eq:defpi}, the Ward identity becomes
obvious, as the $\tau_i$ integrands turn into total derivative upon
contraction with $k_i$,
\begin{equation}
\int_0^T d\tau_i  k_{i,\mu} \dot{x}_{i}^{\mu}e^{ik_ix_{i}} = -i \int_0^T d\tau_i
\frac{d}{d\tau_i} e^{i k_i x_i} =0,\label{eq:zero}
\end{equation}
where the last equality holds as the worldlines are closed,
$x_i(0)=x_i(T)$. However, in any discretized numerical approach, this identity
is difficult to realize as there is no Leibniz rule for latticized
derivatives. Nevertheless, the Ward identity can be exactly maintained in the
numerical worldline algorithm due to the following observation. Consider the
parameter integral combination $I^{\mu_{1}\mu_{2}}$ occurring in \Eqref{eq:defpi},
%
%
\begin{eqnarray}
I^{\mu_{1}\mu_{2}}& = &\int\limits_{0}^{T}d\tau_{1}\int\limits_{0}^{T}d\tau_{2}
~\left(\dot{x}_{1}^{\mu_{1}}e^{ik_1x_{1}}\dot{x}_{2}^{\mu_{2}}e^{-ik_2x_{2}} 
\right)
\end{eqnarray}
\begin{eqnarray}
\phantom{I^{\mu_{1}\mu_{2}}}
&=&\int\limits_{0}^{T}d\tau_{1}\int\limits_{0}^{T}d\tau_{2}~
\left(\dot{x}_{1}^{\mu_{1}}  - 
\frac{k_1^{\mu_{1}}[k_1\dot{x}_{1}]}{k_1^2}\right)
e^{ik_1x_{1}} \nonumber\\
&&\qquad \times\left(\dot{x}_{2}^{\mu_{2}} 
- \frac{k_2^{\mu_{2}}[k_2\dot{x}_{2}]}{k_2^2}\right)e^{-ik_2x_{2}},\label{eq:parint}
\end{eqnarray}
%
%
The mixed terms arising from an expansion of the last equation vanish in the
continuum by virtue of \Eqref{eq:zero}. The same holds for the product of the
second terms in parentheses. This representation explicitly shows that the
tensor structure of $I^{\mu_{1}\mu_{2}}$ is identical to its contraction with
two corresponding transversal projection operators
\begin{align*}
P_{\text{T}}^{\mu\nu}(k_i) = \delta^{\mu\nu} - \frac{k_i^{\mu}k_i^{\nu}}{k_i^2},
\end{align*}
a property which also holds for the local polarization tensor,
\begin{equation}
\pi^{\mu\nu}[k_1,k_2,\xcm;A]= P_{\text{T}}^{\mu\kappa}(k_1)
\pi_{\kappa\lambda}[k_1,k_2,\xcm;A] P_{\text{T}}^{\nu\lambda}(k_2),
\end{equation}
by virtue of which the polarization tensor satisfies the Ward identity
manifestly. In other words, using the identity \eqref{eq:parint}, the
local polarization tensor yields a form
\begin{widetext}

\begin{align}
\pi^{\mu\nu}[k_1,k_2,\xcm;A]
	=& (-ie)^2\int\limits_{0}^{\infty}\frac{dT}{T}\frac{e^{-m^2T}}{(4\pi
          T)^{D/2}}
        \int\limits_{x(0) = x(T),
          \text{CM}}\!\!\!\!\!\!\! 
        \mathcal{D}x~e^{-\int_{0}^{T}d\tau~\frac{\dot{x}^2}{4}}\nonumber  \\
        & \times \left\{\int\limits_{0}^{T}d\tau_{1}\int	
          \limits_{0}^{T}d\tau_	{2}
          \left(\dot{x}_{1}^{\mu_{1}} 
          - \frac{k_1^{\mu_{1}}[k_1\dot{x}_{1}]}{k_1^2}\right)
        e^{ik_1x_{1}}\left(\dot{x}_{2}^{\mu_{2}}  
          - \frac{k_2^{\mu_{2}}[k_2\dot{x}_{2}]}{k_2^2}\right)
        e^{-ik_2x_{2}} e^{-ie\oint dx A} \right\},\label{eq:defpi2a}
\end{align}

\end{widetext}
which satisfies the Ward identity manifestly also upon discretization of the
worldline on a propertime lattice. This discretization can now proceed in the
standard way \cite{Gies:2001zp}. First, we rescale the worldlines,
\begin{equation}
x(\tau) = \sqrt{T} y(t), \quad, \tau=T t,\label{eq:rescale}
\end{equation}
such that the kinetic term
\begin{equation}
\exp \left( -\frac{1}{4} \int_0^T d\tau \dot{x}(\tau)^2 \right)\to
\exp \left( -\frac{1}{4} \int_0^1 dt \dot{y}(t)^2 \right), 
\end{equation}
serving as the probability distribution of the Monte Carlo configurations,
becomes independent of the propertime $T$. Each worldline $y(t)$ is then
represented by a set of $N$ points per loop (ppl), $y_i =y(t_i)$, where
$t_i=i/N$ and $i=1,\dots, N$. The worldline path integral then turns into an
expectation value with respect to an ensemble of $\xcm$ centered worldlines
$\{ y_\ell\}$, where $\ell=1, \dots, \nL$ and $\nL$ denotes the number of
worldlines in the ensemble,
\begin{equation}
        \int\limits_{x(0) = x(T),
          \text{CM}}\!\!\!\!\!\!\! 
        \mathcal{D}x~e^{-\int_{0}^{T}d\tau~\frac{\dot{x}^2}{4}} (\dots)
= \langle(\dots )\rangle.
\end{equation}
The expectation value is normalized to $\langle 1\rangle =1$. There are
powerful algorithms available to generate the ensemble ab initio
\cite{Gies:2003cv,Gies:2005sb}. In the present work, we use the $v$ loop algorithm
\cite{Gies:2003cv}.  Together with the rescaling \eqref{eq:rescale}, the local
polarization tensor then reads
%
%
%
\begin{eqnarray}
&&\pi^{\mu\nu}[k_1,k_2,\xcm;A]\label{eq:defpi2}\\
&&= \frac{(-ie)^2}{(4\pi)^{D/2}}
\int\limits_{0}^{\infty}\frac{dT}{T^{\frac{D}{2}}} e^{-m^2T}
P_{\text{T}}^{\mu\kappa}(k_1)P_{\text{T}}^{\nu\lambda}(k_2) \nonumber\\
&&\quad\times\left\langle\int\limits_{0}^{1}\!\!dt_{1}\!\int	
  \limits_{0}^{1}\!\! dt_{2}
  \dot{y}_{1,\kappa}  e^{i\sqrt{T}k_1y_{1}}
  \dot{y}_{2,\lambda}
  e^{-i\sqrt{T}k_2y_{2}} e^{-ie\sqrt{T}\oint dy A} \right\rangle. \nonumber
\end{eqnarray}
%
%
This representation of the local polarization tensor serves as the master
formula for the construction of our algorithm and its application in the
following sections.

\section{Benchmark tests}
\label{sec:III}

\subsection{Vacuum polarization tensor at zero field}

At $A_\mu=0$, any dependence on the spacetime coordinate drops
out. Homogeneity ensures that the local polarization tensor obeys 4-momentum
conservation, such that \Eqref{eq:defPi} applies. The worldline expression of
the unrenormalized polarization tensor then reads
%
%
\begin{eqnarray}
\Pi^{\mu\nu}[k] 
& =& \frac{(-ie)^2}{(4\pi)^{D/2}} 
\int\limits_{0}^{\infty}\frac{dT}{{T^{D/2}}}{e^{-m^2T}}
P_{\text{T}}^{\mu\kappa}(k)P_{\text{T}}^{\nu\lambda}(k)\label{eq:Pivac} \\
&&\times \left\langle\int\limits_{0}^{1}d\tau_{1}\int\limits_{0}^{1}d\tau_{2}~
            \dot{y}_{1,\kappa}  e^{i\sqrt{T}ky_{1}}
          \dot{y}_{2,\lambda}
        e^{-i\sqrt{T}ky_{2}} \right\rangle. \nonumber
\end{eqnarray}
%
%
For a benchmark test in $D=4$ spacetime dimensions, we need to renormalize
\Eqref{eq:Pivac} and then perform a comparison with the analytically
well-known result for scalar QED \cite{Schubert:2000yt}
%
%
\begin{eqnarray}
\Pi^{\mu\nu}(k)&=&-\frac{e^2}{(4\pi)^2} k^2 P_{\text{T}}^{\mu\nu}(k) \int_0^\infty
\frac{dT}{T} e^{-m^2T}\label{eq:PivacA}\\
&&\quad \times \left\{ \int_0^1 dt e^{-k^2 T t(1-t)} (1-2t)^2 -
  \frac{1}{3} \right\},\nonumber 
\end{eqnarray}
%
%
where the last term in the curly brackets corresponds to the counter-term from
charge renormalization.

Renormalization is not only an important conceptual issue, but also needs to
be taken care of as a matter of practice, as it is advisable to perform the
numerics only for finite renormalized quantities. As the UV divergencies
associated with \Eqref{eq:Pivac} occur for small propertimes $T$, the
divergencies can be analyzed and taken care of on the level of the propertime
integrand which is always finite. Only after the counter terms are subtracted,
we perform the propertime integration which then is perfectly finite. Even
though propertime regularization is most natural for our formalism, our result
can be connected to any other scheme, as the counter terms are known
analytically from the small propertime expansion of the worldline integral.

Still the subtraction is not trivial, as subtracting the analytically known
counter term $(1/3)$ from the numerically evaluated propertime integrand in
\Eqref{eq:Pivac} would lead to an indefinite result with infinitely large
error bars. This is because the small propertime behavior of the numerical
worldline expression is equal to $(1/3)$ only within numerical precision.

The solution to this problem has been provided by \cite{Gies:2001zp} for
effective action computations: the existence of the heat-kernel (small
propertime) expansion of the worldline expectation value at least in an
asymptotic sense allows us to fit the numerically obtained worldline
expression to a polynomial in $T$. Subtracting the constant piece ($=(1/3)$
within numerical errors) then corresponds to charge renormalization. For
improving the stability of the numerical result, it is advisable to fit to a
higher-order polynomial (the coefficients of which can also easily be worked
out analytically with the heat-kernel expansion). The propertime integrand of
$\pi^{\mu\nu}$ is then evaluated with the pure numerical result for
values of $T$ larger than a scale $T_{\text{DG}}$ and with the fit for
$T<T_{\text{DG}}$. The scale $T_{\text{DG}}$ is dynamically generated from the
condition that both expressions for the propertime integrand should have error
bars with the same size at $T_{\text{DG}}$.

\begin{figure}[t]
\includegraphics[scale = 0.23]{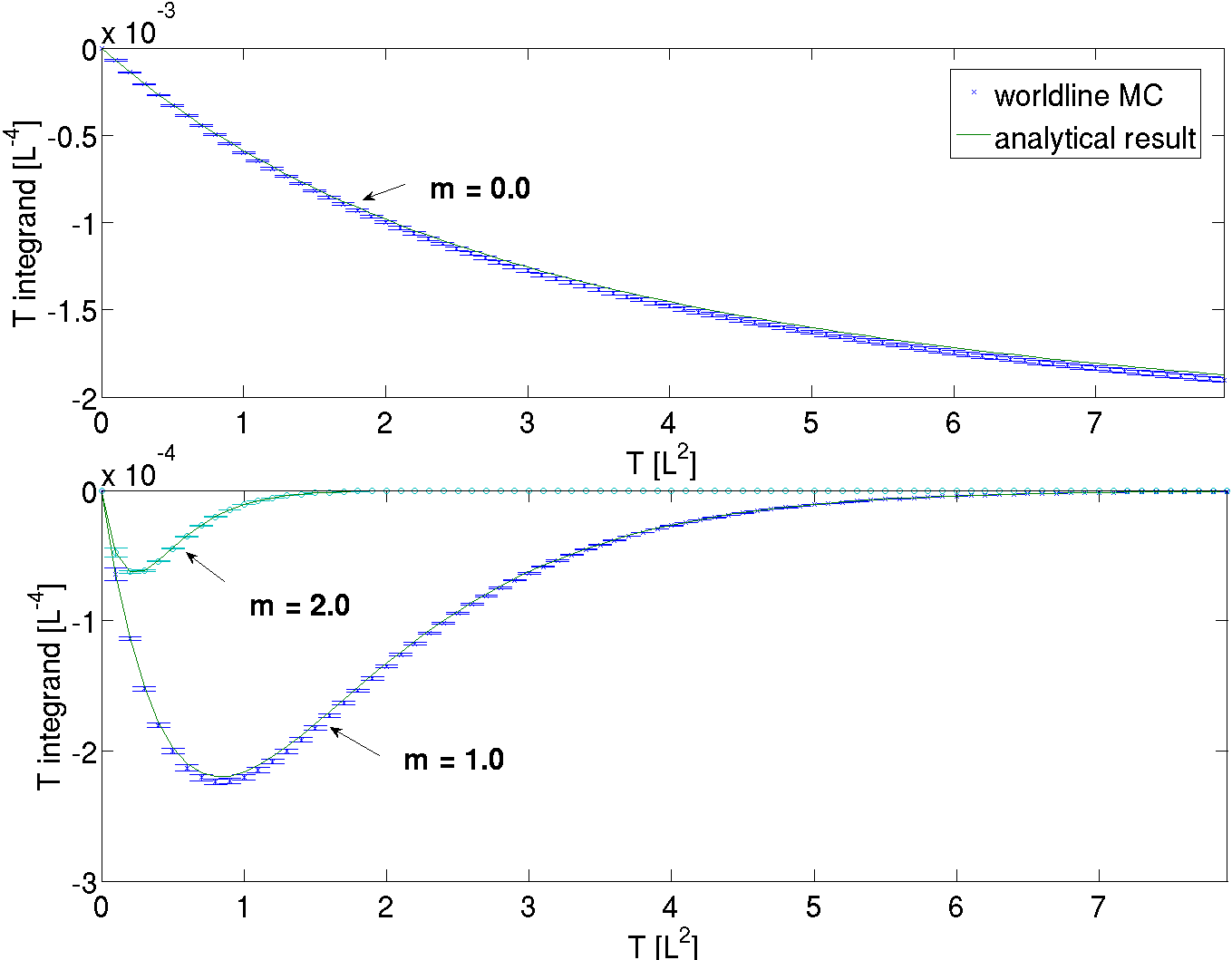}
\caption{Comparison of the analytical and numerical results of the propertime
  integrands of Eqs. \eqref{eq:Pivac} and \eqref{eq:PivacA} for
  $k_\mu=(1,1,1,1)$ (in arbitrary inverse length units $L^{-1}$ and Lorentz indices chosen to point into the 11-direction
  for different values of the electron mass: $m=0$ (upper panel) and $m=1$ and
  $m=2$ (lower panel). In all cases, we have used the same ensemble of random
  worldlines. A test of the algorithm shows that an ensemble with
  $N_{\text{ppl}}=1000$, $\nL=40000$ gives us acceptable results with respect to
  both the calculation time and the numerical errors.}
\label{picture_density_compare}
\end{figure}

In Fig.~\ref{picture_density_compare}, the numerical and analytical results of
the propertime integrand evaluated in Euclidean space for a momentum vector
$k_\mu= (1,1,1,1)$ and Lorentz indices chosen in the 11-direction is
shown. Here, the units are chosen such that each momentum component $k_\mu$ in
arbitrary inverse length units $L^{-1}$ has the value $k_\mu=1$ for $\mu=1,2,3,4$. In the
upper panel, we have set the electron mass $m=0$ whereas the lower panel
depicts the cases $m=1$ and $m=2$ (using the same arbitrary length units). The
latter show a characteristic exponential drop-off for large propertimes
arising from the $e^{-m^2T}$ factor. In all cases, the numerical results
represent a very satisfactory approximation to the analytical results
\cite{Schubert:2000yt}.

\begin{figure*}[t] 
\includegraphics[scale = 0.45]{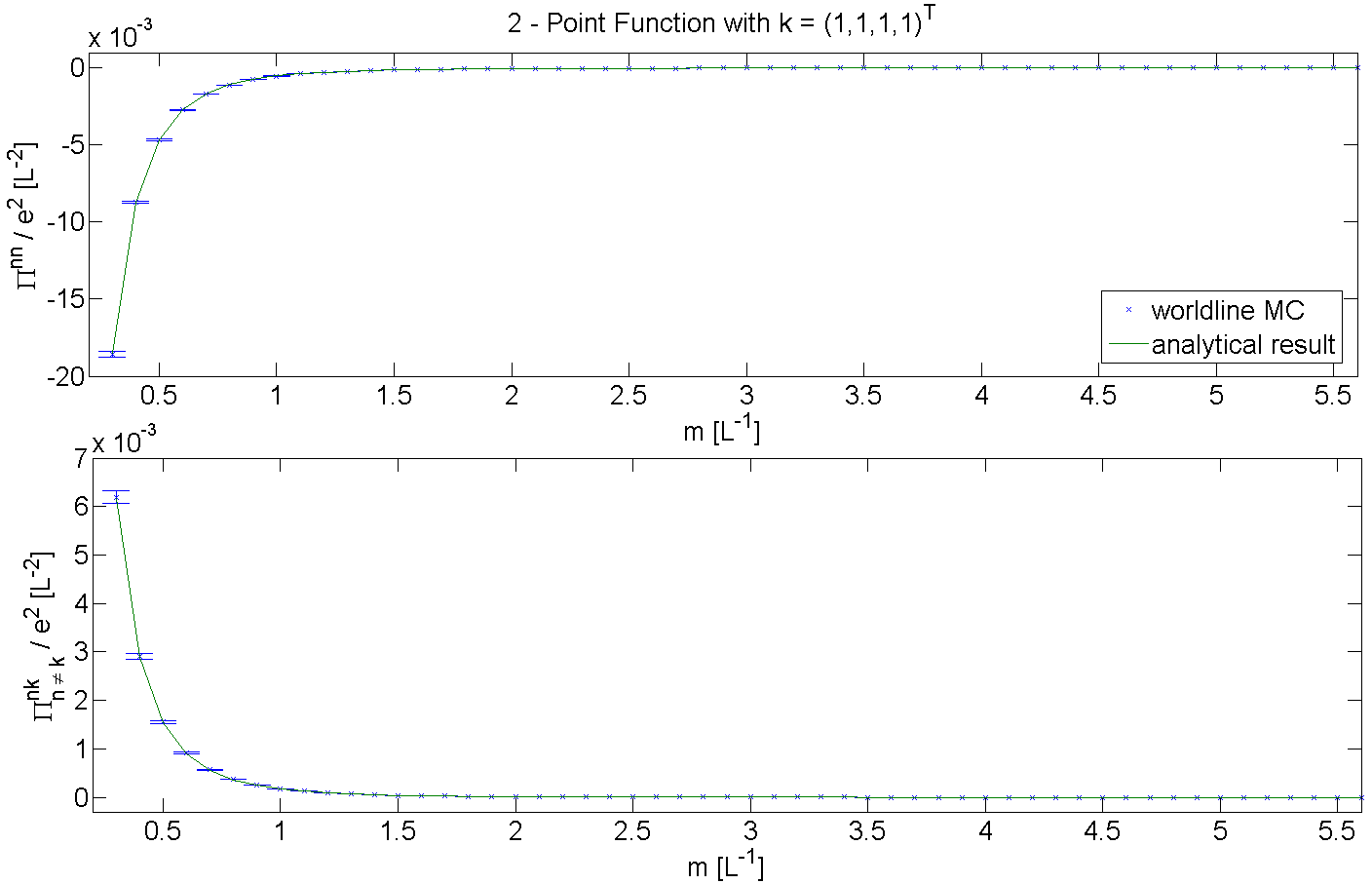}
\caption{Comparison of the analytical and numerical results for the full
  polarization tensor $\Pi^{\mu\nu}$ in scalar QED as a function of the
  electron mass $m$. The upper panel depicts a
  generic result for  the diagonal elements ($\mu = \nu$). The lower panel
  shows the same for off-diagonal elements. }
\label{picture_vacuum_compare}
\end{figure*}

In Fig.~\ref{picture_vacuum_compare}, the diagonal (upper panel) and
off-diagonal (lower panel) components of the vacuum polarization tensor in
Euclidean space are shown as a function of the mass parameter $m$ again for
the case $k_\mu= (1,1,1,1)$. The good agreement between analytical and
numerical results for a wide range of mass values demonstrates that our method
is capable of computing perturbative correlation functions with worldline
numerics.


As described above, we have performed the renormalization by polynomially
fitting the propertime integrand and successively subtracting the counterterm
corresponding to charge renormalization. More precisely, the term in curly
brackets in \Eqref{eq:PivacA} is fitted, for instance, to 
\begin{align}
P =  bT+ cT^2+dT^3,
\end{align}
or higher-order polynomials. The algorithm can be stabilized by inserting the
analytically known coefficients from the heat-kernel expansion, $b=-k^2/30$,
$c=k^4/420$, $d=-k^6/7560$. Similar techniques and knowledge of the
heat-kernel expansion can as well be employed in the case of nonvanishing
electromagnetic fields.

\section{Polarization tensor in a homogeneous magnetic field}
\label{sec:IV}

As is obvious in the worldline approach, the generalization to nonvanishing
background fields is straightforward, by inserting the Wegner-Wilson loop
\begin{align*}
 e^{-ie\int_{0}^{T}d\tau~\dot{x}_{\mu} A^{\mu}(x(\tau))}.
\end{align*}
into the worldline average in \Eqref{eq:Pivac}, cf. \Eqref{eq:defpi2}, with
$x(\tau) =x_{\text{CM}}+\sqrt{T} y(t)$, $t=\tau/T$. As another test of our
method, we compute the vacuum refractive indices arising from fluctuations in
a homogeneous magnetic field. Due to homogeneity, the 4-momentum of the photon
is conserved, implying $k_1+k_2=0$, such that we can directly study the
polarization tensor. Writing the gauge potential in the form
 \begin{align*}
A^{\mu} = \frac{1}{2}F^{\mu\nu}\dot{x}_{\nu} = \frac{1}{2}B\Lambda^{\mu\nu}\dot{x}_{\nu}
\end{align*} 
with the magnetic field strength $B$ and a dimensionless 4-dimensional
tensor $\Lambda^{\mu\nu}$. Choosing the magnetic field to point into the
$\bold{e}_{1}$ direction, $\Lambda^{\mu\nu}$ in Euclidean as well as in
Minkowski space reads:\footnote{In the case of a Minkowskian electric field,
  imaginary Euclidean components would have to be inserted into the worldline
  integrals, see \cite{Gies:2005bz}.}
 \begin{align*}
\Lambda = \left(
\begin{matrix}
0&0&0&0\\
0&0&1&0\\
0&-1&0&0\\
0&0&0&0
\end{matrix}
\right).
\end{align*}
The refractive indices are the inverse of the phase velocities of photons
propagating in a magnetized quantum vacuum. As the magnetic field
distinguishes a direction in space, the magnetized quantum vacuum is
birefringent like a uniaxial crystal, featuring two polarization dependent
phase velocities,
\begin{align}\label{equation_phasevelocity}
v_{\| / \perp}^2 = 1 - \frac{\Pi_{\|,\perp}}{\bold{k}^2} = (1- \Delta v_{\|/\perp})^2,
\end{align}
where $\Pi_{\|,\perp}$ are the nontrivial eigenvalues of the  eigenmodes
$\epsilon_\mu$ of
the polarization tensor \cite{Dittrich:2000zu}, $\Pi^{\mu\nu}
\epsilon_{\mu,\|/\perp}=\Pi_{\|/\perp} \epsilon_{\mu, \|/\perp}$ satisfying
the Minkowski-space photon dispersion relation
\begin{equation}
k^2 + \Pi_{\|/\perp}=0. \label{eq:dispersionrelation}
\end{equation}
Equations \eqref{equation_phasevelocity} and \eqref{eq:dispersionrelation} are actually
identical as the phase velocity is defined as $v= \omega/|\mathbf{k}|$. Here,
we use the Minkowski metric $g=(-,+,+,+)$ and parameterize the Minkowskian
momentum as $(k^\mu)_{\text{M}} = (\omega,|\mathbf{k}|)$.  For
scalar QED, the velocity shifts, $\Delta v_{\|/\perp}=1-v_{\|,\perp}$ in the
weak-field limit yield \cite{Ahlers:2006iz} 
\begin{align}
\Delta v_{\| / \perp} =a_{\|/\perp}
\frac{\alpha}{(4\pi)}\frac{(eB)^2}{m^4} \sin^2\theta, \quad a_{\|/\perp}= \left\{
\begin{matrix}
\frac{1}{90} \\\\
\frac{7}{90}
\end{matrix}\right\}.
\label{eq:velshift}
\end{align}
Here $\theta$ denotes the angle spanned by the magnetic field $\mathbf{B}$ and
the propagation direction $\mathbf{k}$. For weak fields, the $\|$ mode is
polarized in the plane spanned by $\mathbf{B}$ and $\mathbf{k}$, whereas the
$\perp$ mode is polarized orthogonal to this plane; more explicitly,
$\epsilon_\|\sim\widetilde{\Lambda} k$ and $\epsilon_\perp\sim\Lambda k$, with
$\widetilde\Lambda$ denoting the dual field strength matrix.

Even though the worldline numerical formalism is set up in Euclidean space, as
the Monte Carlo procedure requires a positive action, extracting these light
propagation properties obviously requires a transition to Minkowski space. In
particular, we have to insert a Minkowski-valued 4-momentum vector into the
worldline average to have access to real light-cone properties. We do so by
choosing the Euclidean 4-momentum vector as $k_\mu=(i \omega, \mathbf{k})$.
As illustrated in the Appendix, the algorithm remains stable at moderate
frequencies, even though larger fluctuations require typically two orders of
magnitude more statistics than typical Euclidean computations.

\begin{figure}[t]
\includegraphics[scale = 0.23]{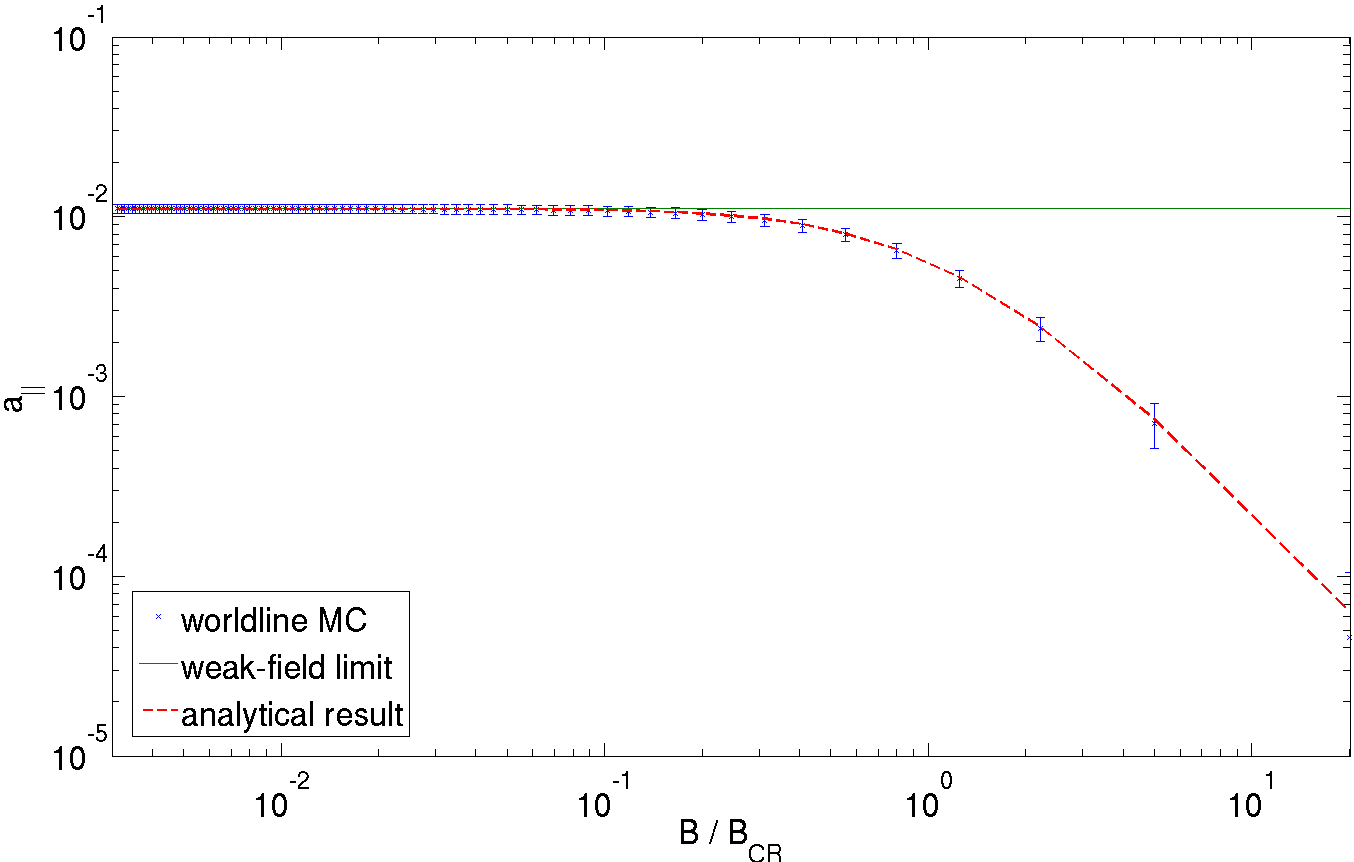}
\caption{Benchmark test: velocity shift prefactor $a_\|$ in a homogeneous
  magnetic field for the $\|$ mode at orthogonal incident $\theta=\pi/2$ and
  for $\omega=0.1m$. For weak fields, $a_\|$ approaches the analytical result
  $1 / 90$ ($\approx 0.01111$), see \Eqref{eq:velshift}. Also for larger
  fields, $a_\|$ follows the analytically known nonperturbative result
  \cite{Ahlers:2006iz}.}
\label{picture_convergence_benchmark}
\end{figure}

In Fig.~\ref{picture_convergence_benchmark}, we compare our worldline
Monte-Carlo results with those of the analytically known velocity shifts
\cite{Ahlers:2006iz}  over a wider range of magnetic field strength (also
exceeding the simple weak field limit). This benchmark test is performed for
orthogonal incident $\theta=\pi/2$ and for the $\|$ mode at a frequency
$\omega=0.1 m$. The numerical results approach the velocity shift in the
weak-field limit with very good accuracy and also yield reliable results for
larger field strengths. 

\section{Polarization tensor in a  spatially inhomogeneous magnetic field}
\label{sec:inhom}

Let us now explore new vacuum polarization effects in inhomogeneous fields,
revealing the nonlocal nature of fluctuation-induced processes. For this, we
use a magnetic background pointing into, say, $\mathbf{e}_1$ direction,
consisting of a constant magnetic field $\overline{B}$ superimposed with a sinusoidal
oscillation varying in $\mathbf{e}_3$ direction with amplitude $B_1$ and
wavelength $\lambda_B$,
\begin{equation}\label{eq:Binhom}
\mathbf{B}(x_3) = \left[\overline{B} +
  B_{1}\cos\left(\frac{2\pi}{\lambda_{B}}x_{3}\right)\right]\bold{e}_{1}.
\end{equation}

\begin{figure}[t] 
\includegraphics[width = \linewidth]{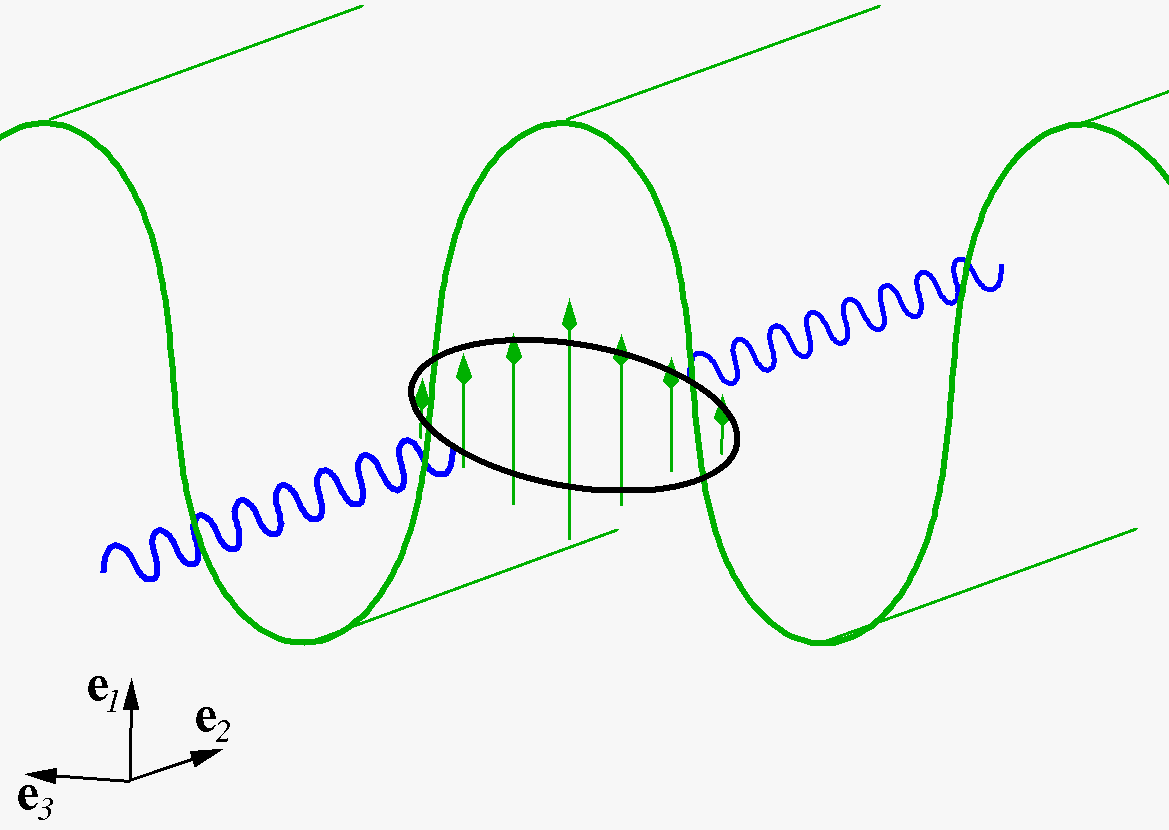} 
\caption{Sketch of the geometry of our configuration with an inhomogeneous
  magnetic field $\text{\textbf{B}}(x_{3})$ (green arrows and corrugated
  surface) with a constant magnetic component $\overline{B}$ in
  $\mathbf{e}_{1}$ direction and a spatial variation along the
  $\mathbf{e}_{3}$ direction with field amplitude $B_{1}$ also pointing into
  the $\mathbf{e}_1$ direction. The photon (blue wiggly line) propagates
  perpendicular to the field in $\mathbf{e}_{2}$ direction. The
  fluctuation-induced interaction between the photon and the magnetic field is
  represented by a Feynman diagram (black ellipse).}
\label{picture_schema_inhomogen}
\end{figure}

A similar electric field has already been used to analyze the role of spatial
inhomogeneities in Schwinger pair production \cite{Gies:2005bz}. 
This field configuration can be viewed as a rough approximation to a realistic
strong and broad laser pulse in standing wave mode superposed with higher harmonics. 

For the
worldline simulation, we use the corresponding gauge potential
\begin{align}\label{eq_Aosc}
A_2 = -\overline{B}x_{3} -
\frac{B_{1}\lambda_{B}}{2\pi}\sin\left(\frac{2\pi}{\lambda_{B}}x_{3}\right), 
\end{align}
which is numerically convenient, as it depends only on one spatial
coordinate. 

As a relevant observable, we compute the local velocity shift $\Delta v(x_3)$
for a photon propagating along the $\mathbf{e}_2$ direction from the local
polarization tensor. The geometry of our configuration is sketched in
Fig.~\ref{picture_schema_inhomogen}. As the magnetic field is homogeneous in
$\mathbf{e}_2$ direction, the photon momentum is conserved $k_1+k_2=0$ for
$\mathbf{k}_1 \sim \mathbf{k}_2\sim \mathbf{e}_2$. Hence, setting $k=k_1=-k_2$
and inserting \eqref{eq_Aosc} into the local polarization tensor
\eqref{eq:defpi2}, we can determine $\pi^{\mu\nu}[k,x_{3, \text{CM}};A]$,
which is diagonalized by the same polarization eigenmodes $\epsilon_{\mu,
  \|/\perp}$ as the in constant-field case. The local phase velocity shifts then
are computed analogous to \Eqref{equation_phasevelocity},
\begin{equation}\label{equation_phasevelocity_in}
v_{\| / \perp}^2(x_{3, \text{CM}}) = 1 - \frac{\pi_{\|,\perp}(x_{3,
    \text{CM}})}{\bold{k}^2} =
 (1- \Delta v_{\|/\perp}(x_{3, \text{CM}}))^2,
\end{equation}
where $\pi_{\|,\perp}(x_{3, \text{CM}})$ are the local eigenvalues of the
polarization tensor.

 \begin{figure}[t]
\includegraphics[scale = 0.21]{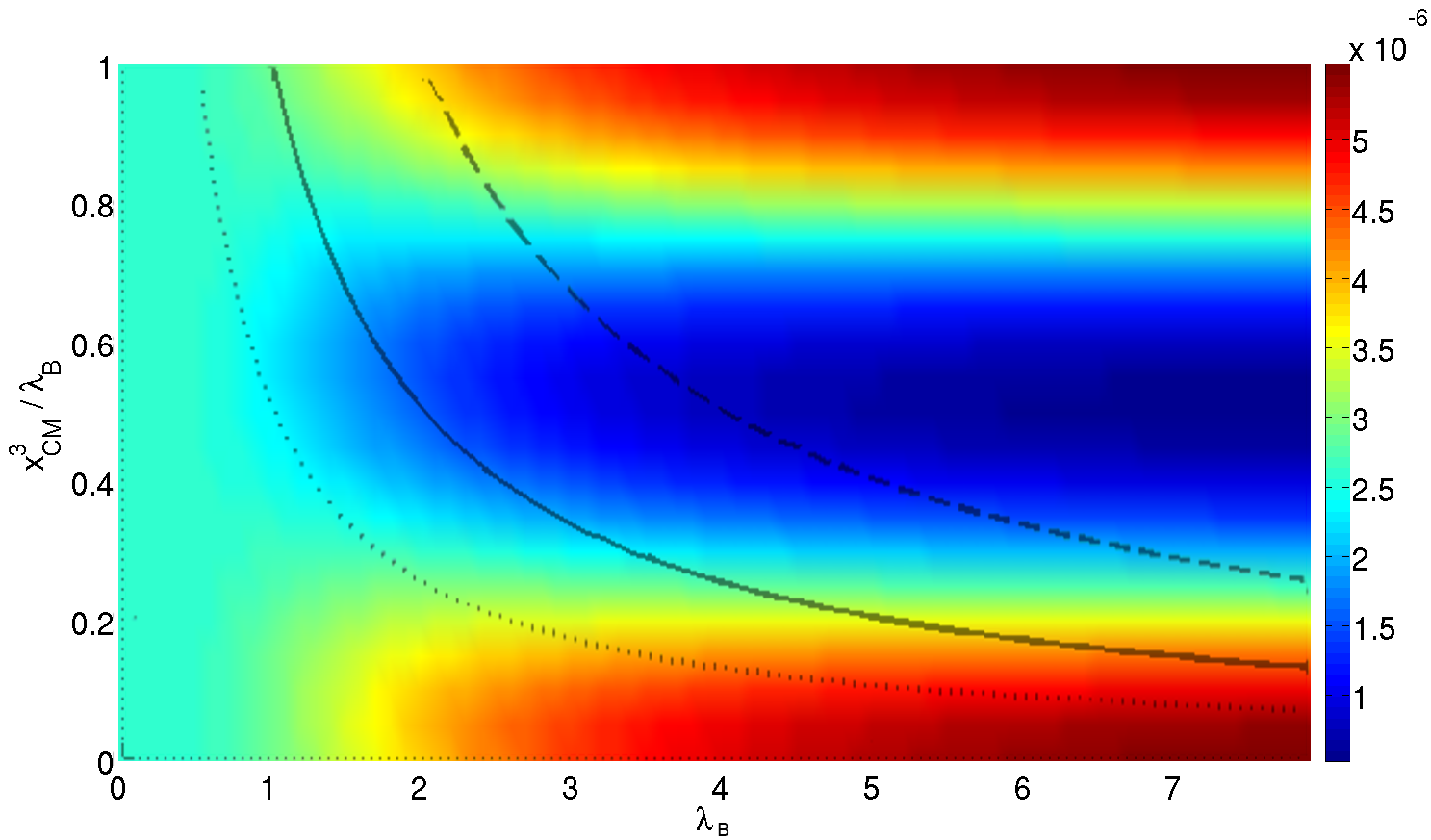} 
\caption{Contour plot of the phase velocity shift $\Delta v$ for an incoming
  $\|$-photon at orthogonal incident with respect to the
  direction of the external field ($\theta = \frac{\pi}{2}$). The
  inhomogeneous field has a sinusoidal variation along the $\mathbf{e}_3$
  direction, cf. \Eqref{eq:Binhom}, with $e\overline{B}=0.2 m^2$ and $B_{1} =
  0.5 \overline{B}$. The phase velocity is plotted horizontally versus the oscillation
  wavelength $\lambda_B$ and vertically versus the normalized position $x_{3,
    \text{CM}}/\lambda_B$ of the photon relative to the oscillation
  phase. Lines of constant absolute position $x_{3, \text{CM}}=0.5,1.0,2.0$ in units of
  mass $m$ for varying wavelength $\lambda_B$ are indicated
  as dotted, solid and dashed lines.
}
\label{picture_map}
\end{figure}

 \begin{figure}[t]
\includegraphics[scale = 0.18]{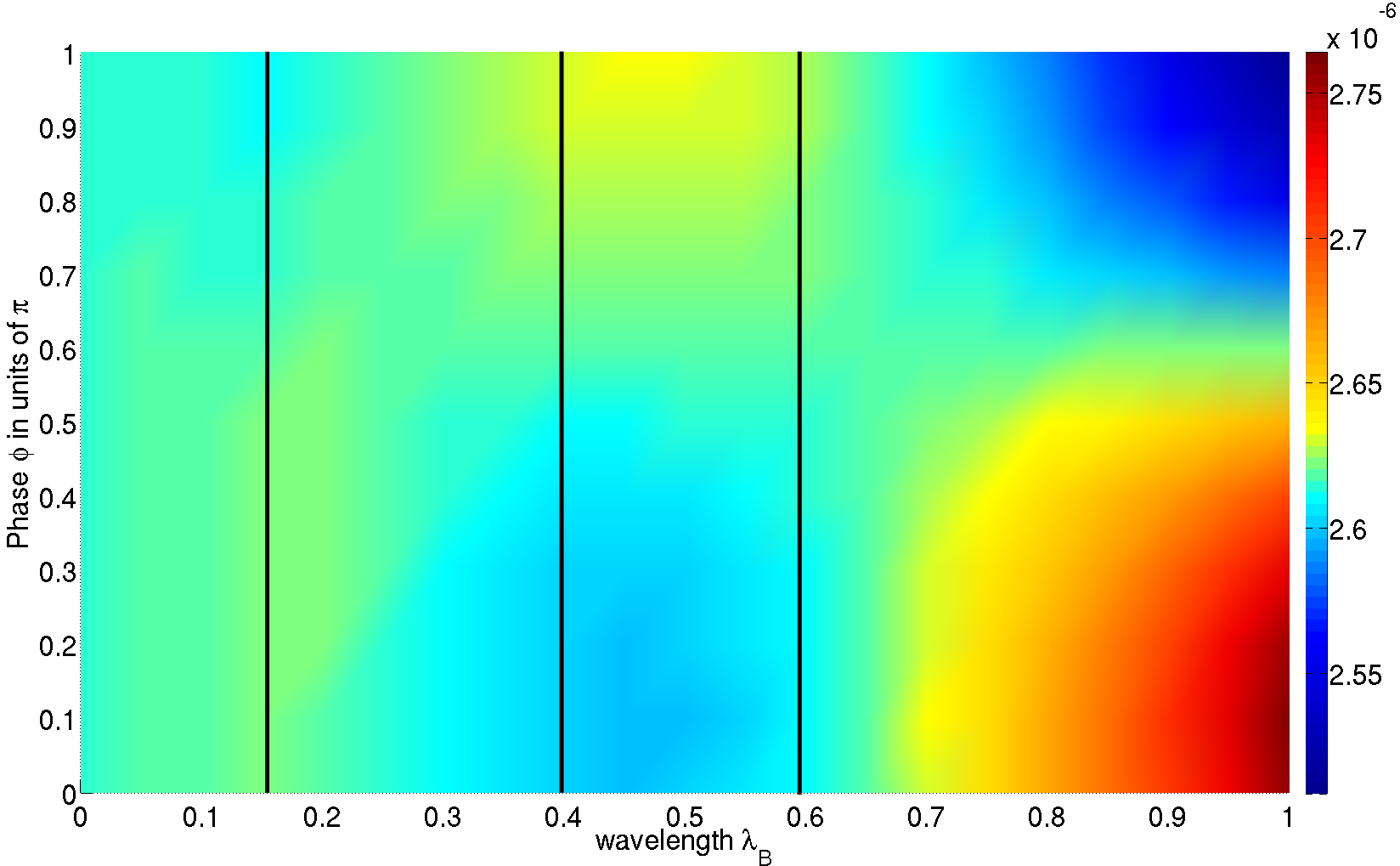} 
\caption{Contour plot with the same physical parameters as in Figure
  \ref{picture_map} but with a higher resolution in the region of small
  $\lambda_{B}$. The $y$ axis is now given in terms of the phase in units of
  $\pi$, $\phi/\pi = 2 x_{\text{CM}}/\lambda_B$. The straight lines are cuts
  at $\lambda = 0.15, 0.4, 0.6$, which we discuss in Figure
  \ref{Inhomogen_Phase}.  }
\label{picture_map_hr}
\end{figure}
In our computation, we use $e\overline{B}=0.2 m^2$ and $B_{1} = 0.5
\overline{B}$. Figure \ref{picture_map} shows a contour plot of the local
velocity shift as a function of the oscillation wavelength $\lambda_B$ and the
normalized coordinate $x_{3, \text{CM}}/\lambda_B$ for the $\|$-mode; the
latter corresponds to the phase inside the oscillation period, $\phi=2\pi
x_{3, \text{CM}}/\lambda_B$. For large $\lambda_B$ when the field becomes slowly varying
with respect to the Compton wavelength $\lambda_B m\gg 1$, the local velocity
shifts approach the constant field values as expected. A
``locally-constant-field'' approximation becomes reliable in this limit. By
contrast, if the two characteristic length scales become similar $\lambda_B
m\simeq 1$ the oscillating structure of the magnetic field starts to become
washed out in the local velocity shift. The propagating photon undergoing
virtual electron-positron loops with an inherent length scale of the Compton
wavelength ``sees'' a field averaged over this length scale. For a rapidly
oscillating field, $\lambda_B m \ll 1$, the field oscillations are completely
washed out and become invisible in the velocity shift. In our present example,
the limiting velocity shift in this rapid-oscillation limit corresponds
precisely to that induced by the background field $\overline B$, $\Delta v
\sim \overline{B}^2$. This is in line with the interpretation that the
averaging arises from the nonlocal nature of the fluctuations. For instance,
naively averaging over the constant-field velocity shift, depending
quadratically on $B$, would give a different (and wrong) averaging result,
$\Delta v \not\sim (\overline{B}^2 + \frac{1}{2} B_1^2)$.

\begin{figure}[t]
\includegraphics[scale = 0.22]{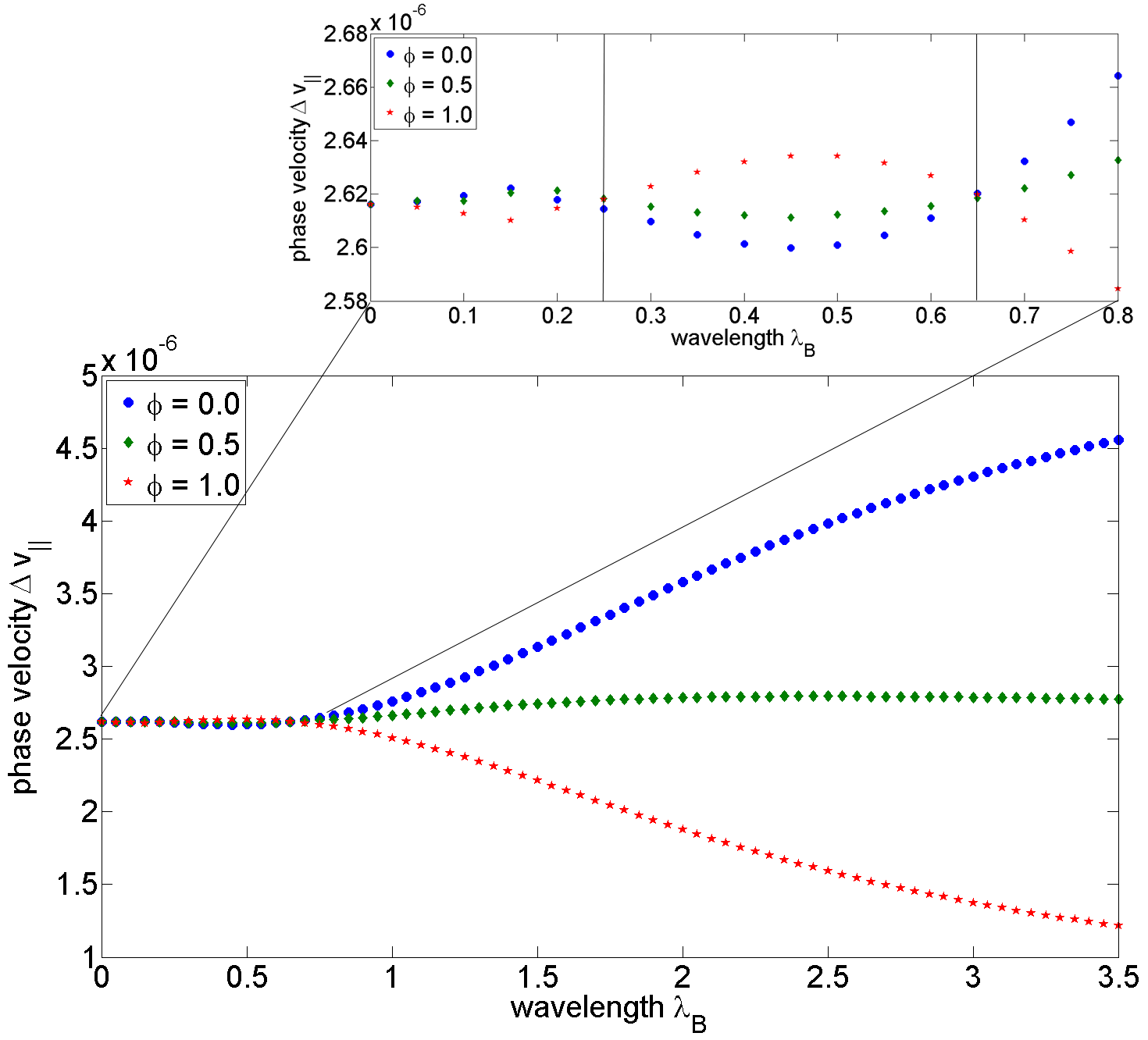} 
\caption{Phase velocity shift $\Delta v_\|$ for different positions in the
  phase of the variation $\phi=2 \pi x_{3, \text{CM}}/\lambda_B$ as a function of the
  variation length $\lambda_B$.  The curves correspond to horizontal slices of
  the contour plot Figs.~\ref{picture_map},\ref{picture_map_hr} at the
  ordinate values $\phi/\pi=2x_{\text{CM}}/\lambda_B=0,0.5,1$. For large
  $\lambda_B$, the velocity shifts approach their constant-field limits with a
  clear ordering from large to small background field from top to bottom (blue
  dots at field maximum to red stars at minimum field). By contrast, this
  ordering is modified at small $\lambda_B$ in various patterns depending on
  the value of $\lambda_B$ (see inlay). The straight lines in the inlay at
  $\lambda_{B} \simeq 0.25, 0.65$ mark the turning points.  }
\label{error}
\end{figure}

An interesting parameter regime occurs at small $\lambda_B$, i.e., for rapidly
varying background field, see Fig.~\ref{picture_map_hr}. In Fig.~\ref{error},
we show horizontal slices of the contour plot Fig.~\ref{picture_map_hr} at the
ordinate values $x_{3, \text{CM}}/\lambda_B=0,0.25,0.5$, corresponding to different
positions in the phase of the variation $\phi/\pi=2 x_{3, \text{CM}}/\lambda_B=0,0.5,1$,
as a function of the variation length $\lambda_B$. For large $\lambda_B$, the
velocity shifts approach their constant-field limits with a clear ordering
from large to small background field from top to bottom. This is in accordance
with expectations from a locally-constant-field approximation becoming
applicable for large $\lambda_B$. 

By contrast, this ordering is lost at small $\lambda_B$, where the curves show
a characteristic oscillation pattern, see inlay of Fig.~\ref{error}. Depending
on the value of $\lambda_B$, the velocity shift in a local minimum of the
field (red stars, $\phi/\pi=1$) can become larger than that in a local maximum
(blue dots, $\phi/\pi=0$), cf. inlay of Fig.~\ref{error} at around
$\lambda_B\simeq 0.65$.
We interpret this
phenomenon as a consequence of the local averaging property of the quantum
fluctuations on scales of the Compton wavelength $m$. In this way, the local
velocity shift in a minimum of the field can receive dominant contributions
from the nearby maxima if they are significantly probed by the quantum
fluctuations on the scale $1/m$. Conversely, the local velocity shift in a
maximum of the field can receive dominant contributions from the nearby
minima. This can lead to an inversion of the hierarchy of the velocity shifts
with respect to the local background field. Our data is compatible with
further oscillations setting in at even smaller values of
$\lambda_B\simeq0.25$, which would correspond to further minima or maxima
entering the local fluctuation average.


\begin{table*}
\begin{tabular}{|c|c|c|c|}
\hline
variation length $\lambda_{B}$&amplitude $A$&phase shift $\phi_0 / \pi$&constant velocity shift $\Delta v_{0}$\\ \hline\hline
0.15&7.36$\cdot10^{-9}\pm$2.5$\cdot10^{-9}$&0.31$\pm$0.30&2.62$\cdot10^{-6}\pm$3.4$\cdot10^{-8}$\\
0.40&1.61$\cdot10^{-8}\pm$4.8$\cdot10^{-8}$&1.09$\pm$5.4$\cdot10^{-1}$&2.62$\cdot10^{-6}\pm$5.1$\cdot10^{-8}$\\
0.60&8.66$\cdot10^{-9}\pm$3.4$\cdot10^{-8}$&1.13$\pm$6.8$\cdot10^{-1}$&2.62$\cdot10^{-6}\pm$5.1$\cdot10^{-8}$\\
2.00&8.52$\cdot10^{-7}\pm$2.1$\cdot10^{-7}$&0.02$\pm$3.0$\cdot10^{-2}$&2.74$\cdot10^{-6}\pm$3.8$\cdot10^{-8}$\\ \hline
\end{tabular}
\caption{Fit parameters deduced from the numerical results of
  Fig.~\ref{Inhomogen_Phase} using a least-squares fit to \Eqref{eq:fit}.
  Starting from large $\lambda_B$ (slowly varying field), the velocity shift
  is in phase with the external field, $\phi_0\simeq 0$. For
  $\lambda_B=0.4,0.6$, we observe a jump of the phase by $\phi_0/\pi\simeq
  1$. For an even smaller variation scale $\lambda_B=0.15$ the phase $\phi_0$
  goes back to values compatible with zero (the deviations from zero arise from the
  comparatively large signal to noise ratio of our data for small
  $\lambda_B$; this is also reflected in the large error bars for the
  amplitude $A$ for smaller $\lambda_B$). }
\label{tab_fitting_results}
\end{table*}

\begin{figure}[h]
\includegraphics[scale = 0.22]{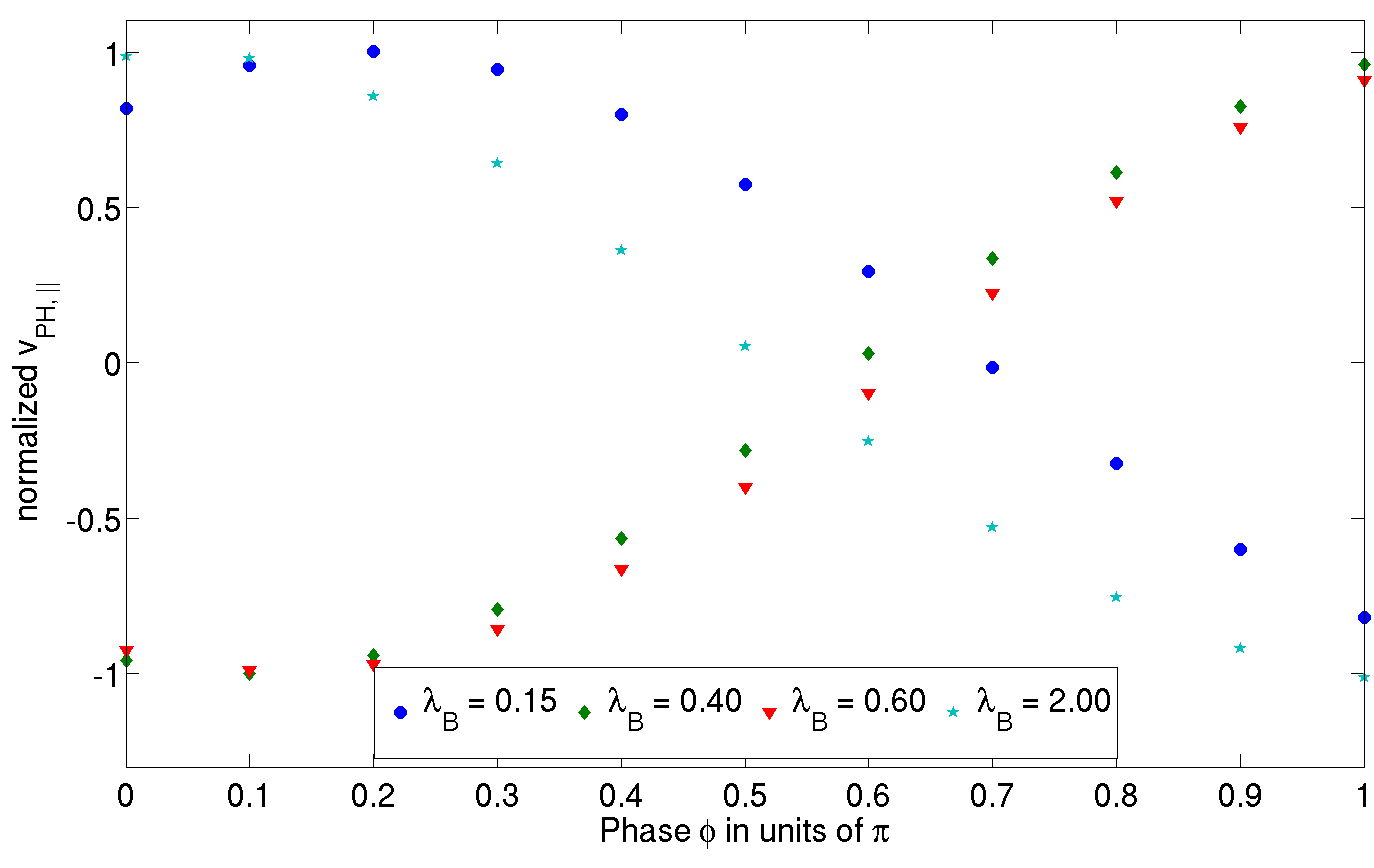}
\caption{Normalized phase velocity shift $\Delta v_\|$ for different field
  inhomogeneities $\lambda_B=0.15,0.40, 0.6,2.0$ as a function of the phase of the
  field variation $\phi/\pi=2  x_{3, \text{CM}}/\lambda_B$. The curves correspond to
  vertical slices of the contour plot Fig.~\ref{picture_map_hr} at the given
  abscissa values for $\lambda_B$.  The picture shows the expected
  $2\pi$-periodicity of the phase velocity with respect to $\phi$. For
  large $\lambda_B=2$ (light blue stars), the velocity shift is in phase with the background
  field $\sim \cos \phi$ as expected from a locally-constant field
  approximation. For $\lambda_B=0.4,0.6$ (green diamongs, red triangles), the
  velocity shift oscillates out of phase $\sim \cos (\phi-\phi_0$ with $\phi_0
  \simeq \pi$. For even smaller variation length $\lambda_{B}=0.15$, the
  oscillation is approximately in phase again $\sim \cos \phi$.  }
\label{Inhomogen_Phase}
\end{figure}

This becomes visible in vertical slices of the contour plot
Fig.~\ref{picture_map_hr} at the abscissa values $\lambda_B=0.15,0.4,0.6$, and
$2$ as a function of the phase of the field variation $\phi=2 \pi x_{3,
  \text{CM}}/\lambda_B$.  In Fig.~\ref{Inhomogen_Phase}, we show the varying
part of the velocity shift normalized by the maximum oscillation amplitude
$A$. We observe the expected $2\pi$-periodicity of the phase velocity with
respect to $\phi$. Most importantly, the oscillation is shifted by an offset
$\phi_0=\pi$ for small wavelengths, e.g., $\lambda_{B}=0.4,0.6$, compared to
larger $\lambda_B=2$ where the velocity shift tends to approach the
locally-constant-field limit. This implies that velocity shift minima occur at
field maxima and vice versa in the range $0.25\lesssim \lambda_B \lesssim
0.65$. For even more rapid field oscillations $\lambda_B\lesssim 0.25$, our
data is compatible with the velocity shift being in phase with the external
field again. This goes hand in hand with our interpretation in terms of
fluctuation averages. The full Monte Carlo data can be parametrized by a
simple fit,
\begin{align}
f(x) = A\cos\left(2\pi \frac{x_{\text{CM}}}{\lambda_B} - \phi_0 \right) +
\Delta v_{0}.
\label{eq:fit}
\end{align} 
The fitting results are given in
Tab.~\ref{tab_fitting_results}.


A similar phenomenon had already been observed for the case of pair production
in inhomogeneous fields \cite{Gies:2005bz}; however, due to the exponential
dependence of pair production on the background field, this averaging
phenomenon was much more pronounced in this case. 
In fact, the amplitude of the oscillations is on the order of our statistical error
bars for
the velocity shift data. However, fitting the data at fixed $\lambda_B$ to
sinusoidal fit functions translates into a correspondingly large error for the
amplitude, but a significantly small error for the phase of the
oscillation, see Tab.\ref{tab_fitting_results}. The size of our statistical
errors can also be estimated from the deviations from symmetry about the
horizontal $x_{3, \text{CM}}/\lambda_B=0.5$ axis in Fig.~\ref{picture_map} or from
anti-symmetry about the horizontal $\phi=0.5$ axis in
Fig.~\ref{picture_map_hr}. This latter anti-symmtry also guarantees that the
exact result for the phase velocity shift at $\phi=0.5$ should not depend on
$\lambda_B$; the slight dependence of the Monte Carlo data for
the $\phi=0.5$ curve (green diamonds) on $\lambda_B$, hence is a measure for
our statistical error.\footnote{Of course, these (anti-)symmetries could be
  implemented explicitly in the Monte Carlo computation by generating
  correspondingly symmetric ensembles.}  

As these velocity shifts correspond to shifts of the local
refractive indices this new phenomenon has a direct consequence for the
self-focussing property of the quantum vacuum  \cite{Kharzeev:2006wg}: in
the locally-constant-field limit (a pure Heisenberg-Euler-type calculation),
the refractive index increases with increasing field strength. This implies
that photons are dragged into local maxima of the field strength. This even
enhances the field strength at local maxima, thus giving rise
to self-focussing properties. Our observation in turn predicts that this
self-focussing naturally terminates on the scale of the Compton wavelength. If
field maxima are self-focussed down to a critical scale $\lambda_{\text{cr,1}}$,
field maxima with nearby minima can become local minima of the velocity shift
(and thus minima of the local refractive index) such that the quantum vacuum
becomes defocussing again. From the inlay of Fig.~\ref{error}, we estimate
this critical scale to be near $\lambda_{\text{cr,1}} \simeq 0.65$ in units of
the Compton wavelength. This critical scale provides a natural limit
to the self-focussing property of the quantum vacuum. As our data is
compatible with an in-phase dependence of the refractive index on the field
inhomogeneities for $\lambda_B\lesssim 0.25$, the quantum vacuum may become
self-focussing again on this shorter variation scale. But this regime is also
expected to terminate at another critical scale $\lambda_{\text{cr,2}}$ where
the velocity shift may run out of phase again. Our data is compatible with 
$\lambda_{\text{cr,2}}\lesssim 0.05$.

\section{Conclusions}
Based on the successful worldline approach to perturbative correlation
functions, we have developed numerical Monte-Carlo techniques for the
computation of the vacuum polarization tensor in inhomogeneous background
fields for scalar QED. These techniques generalize earlier methods which
have been frequently applied to effective action or quantum energy
computations. The new challenge in the case of correlation functions is the
appearance of further scales provided by the incoming and outgoing momenta. 

The stability of the numerical algorithm also originates in the fact that it
satisfies the Ward identity exactly and operates on the level of renormalized
quantities. We have explicitly demonstrated that the algorithm can also be
used to determine correlation functions as a function of Minkowski-valued
momenta and fields, even though stability is expected to become an issue for
increasing Minkowski momenta or dominating electric field components.

We have verified our algorithm with the analytically known cases of the vacuum
polarization tensor for off-shell momenta and the polarization tensor in
homogeneous fields using the magnetically induced light-cone deformations as
an observable. In these cases, the algorithm is capable to reach a precision
on the percent level at moderate numerical cost. 

Furthermore, we have studied for the first time light propagation in a
spatially varying magnetic field. For small variations of the field compared
to the Compton wavelength, the local derivative expansion (or
locally-constant-field approximation) is well applicable as expected, such
that the vacuum polarization tensor quickly approaches the constant-field
limit. 

For rapidly varying fields, the vacuum-magnetic refractive indices can exhibit
a non-monotonic dependence on the local field strength. This new behavior can
geometrically be understood in the worldline picture, as the worldlines and
their spatial extent probes the nonlocal structure of quantum field
theory. Local values of the refractive indices can receive dominant
contributions from nearby maxima or minima of the field strength. This
inherent averaging mechanism induces a smearing and even non-monotonical
features of the refractive indices. For
the properties of light propagation, this can provide a natural limit on the
self-focussing property of the quantum vacuum.

Our present study represents a first step into the largely unknown territory of
quantum correlation functions in inhomogeneous fields. Even though we have
concentrated on the two-point function in the present work, we expect that our
algorithmic strategy can rather straightforwardly be generalized to
higher-order correlation functions. Also a generalization to spinor QED is in
principle straightforward and merely requires the inclusion of the worldline
spin factor involving a numerically moderately expensive path-ordering
prescription.

\acknowledgments The authors thank B. Doebrich, G.V. Dunne, F. Karbstein, S.~P.~Kim,
C. Schubert, and A. Wipf for helpful discussions and acknowledge {support by
  the DFG under grants Gi~328/3-2, and SFB/TR18. HG thanks the DFG for support
  through grant Gi~328/5-1 (Heisenberg program). LR is grateful to the
  Carl-Zeiss Stiftung for financial support (PhD fellowship).

\appendix

\section{Validity control of the numerical algorithm}

The worldline Monte Carlo method has proven its efficiency and accuracy in many
examples in the context of effective action and quantum energy
computations. Generically, the convergence is very satisfactory and scales
with a typical Monte Carlo $1/\sqrt{\nL}$ dependence, where $\nL$ is the
number of configurations, i.e., worldlines in this case. Precision with an
error on the 1\% level can be achieved with moderate numerical cost. In the
present case, it is worthwhile to critically re-examine the quality and
efficiency of the algorithm, as the calculation of correlation functions goes
along with further technical requirements. Most prominently, the physical
observables need to be computed with Minkowski-valued momenta which
technically is a potential source of numerical instability. This is because
the Euclidean phase factors $\sim \exp(i k_\mu x_\mu)$ receive real
exponential contributions for Minkowskian momenta $k_0 \to i \omega$. Also, the
frequency and spatial momentum dependence introduces further scales which can
interfere with the scale of inhomogeneity. 

All error estimates in this work are based on the Jackknife method. We have
checked explicitly, that this error estimate using the same random number seed
yields results  equivalent to an error estimate derived from a set of
ensembles created with different random number seeds. 

In Fig.~\ref{picture_Error_Size}, we compare the relative error for the
velocity shift $\Delta v_\|$ in percent as a function of the number of worldline
configurations $\nL$ for various parameters. The smallest error is observed
for a purely Euclidean constant field calculation (green dots). This type of
calculation is closest to conventional  effective action and quantum energy
computations; however, in the present case it does not describe a physical
observable. The corresponding physical Minkowskian calculation for the
constant-field case (blue diamonds) shows an error increase of roughly an
order of magnitude. This implies that an error on the level of a typical
Euclidean calculation requires two orders of magnitude more statistics.

\begin{figure}[t]
\includegraphics[scale = 0.22]{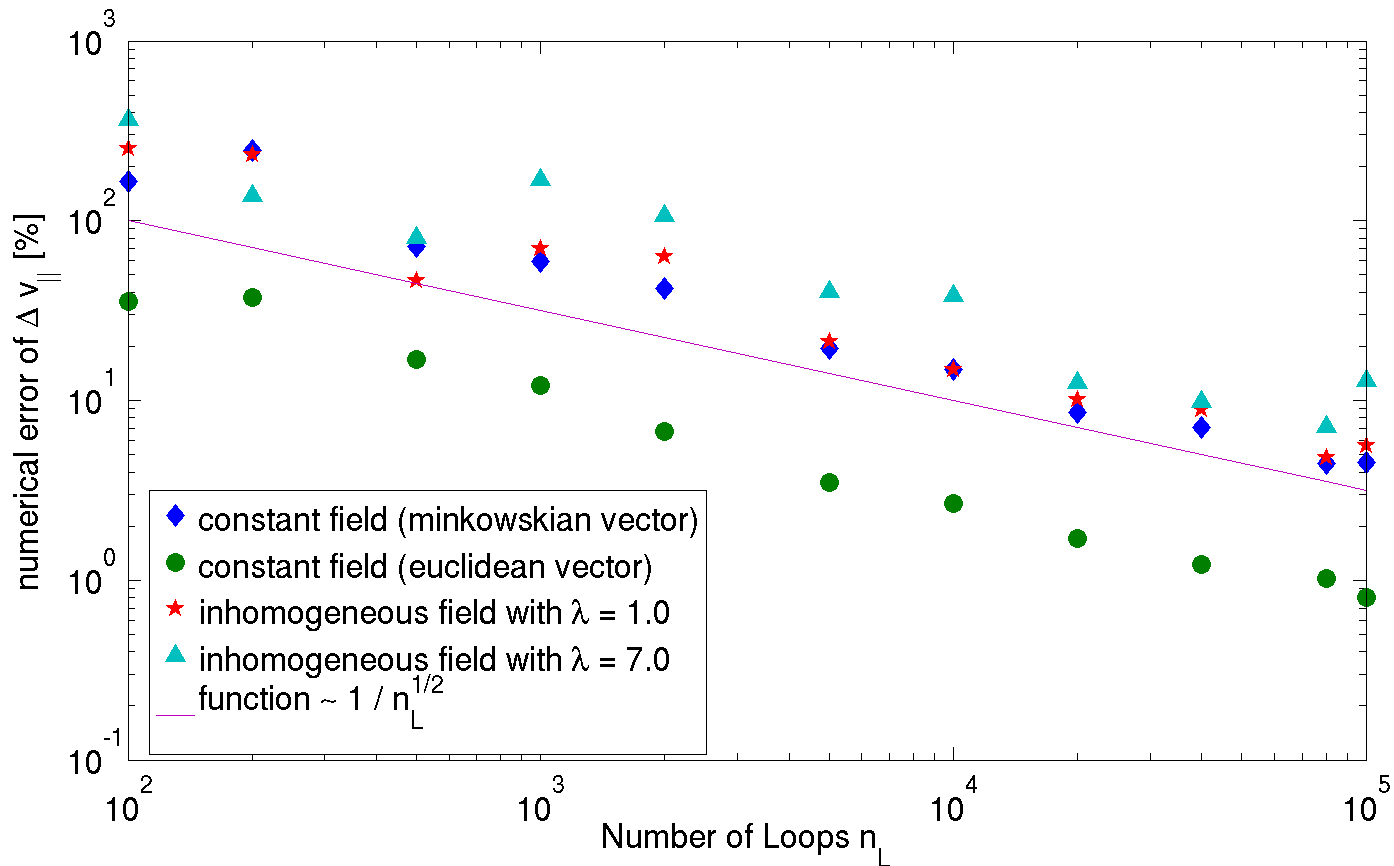}
\caption{Relative error (jackknife estimate) for the velocity shift $\Delta v$
  in percent as a function of $\nL$ for a Euclidean vs. Minkowskian calculation
  including its dependence on the field inhomogeneity. We have used a
  worldline ensemble with $N=1000$ points per loop. The plot shows a
  comparison between the Euclidean constant-field case with
  $e\overline{B}/m^2=0.2$ and a
  Minkowskian calculation for the same constant-field case (blue diamonds).
  Two Minkowskian calculations for the inhomogeneous field of
  Sect.~\ref{sec:inhom} with wavelengths $\lambda_B = 1.0$ (red stars) and
  $\lambda_B = 7.0$ (light blue triangles) are also shown. Because of the
  stronger fluctuations in the Minkowskian case, all Minkowskian calculations
  show larger relative errors. The dependence of the error on the background
  inhomogeneity is comparatively minor. For illustration, also a line $\sim
  1/\sqrt{\nL}$ is shown reflecting the expected error depletion for Monte Carlo
  calculations.}
\label{picture_Error_Size}
\end{figure}
 
The interference of the error with the external scales is also illustrated in
Fig.~\ref{picture_Error_Size}, where we determine the relative error for the
velocity shift $\Delta v$ in percent as a function of $\nL$ and its dependence
on the field inhomogeneity for two different values $\lambda_B$ parameterizing
the field inhomogeneity. Even though we observe some dependence on the field
inhomogeneity leading to slightly larger errors, the main effect on the error
clearly arises from the necessity to perform Minkowski-valued calculations.

Fig.~\ref{picture_Error_Size} also depicts a straight line exhibiting a
$\sim1/\sqrt{\nL}$ dependence in this double-log plot. This indicates that the
numerical error decreases with increasing number of worldlines $\nL$ as
$\sim1/\sqrt{\nL}$, as expected. Even for these perturbatively small
Minkowski-valued quantities, errors below the 10\% level are achievable at
managable numerical cost.



\begin{thebibliography}{99}

\bibitem{Toll:1952}
J.~S.~Toll,
Ph.D. thesis, Princeton Univ., 1952 (unpublished).


\bibitem{Baier}
R.~Baier and P.~Breitenlohner, 
Act.~Phys.~Austriaca {\bf 25}, 212 (1967); 
Nuov.~Cim.~B {\bf 47} 117 (1967).

\bibitem{BialynickaBirula:1970vy}
  Z.~Bialynicka-Birula and I.~Bialynicki-Birula,
  Phys.\ Rev.\  D {\bf 2}, 2341 (1970).


\bibitem{Adler:1971wn}
  S.~L.~Adler,
  Annals Phys.\  {\bf 67}, 599 (1971).

\bibitem{Tsai:1975iz}
  W.~y.~Tsai and T.~Erber,
  %
  Phys.\ Rev.\ D {\bf 12}, 1132, (1975).


\bibitem{Gies:1999vb}
  H.~Gies,
  Phys.\ Rev.\ D {\bf 61},  085021  (2000).


\bibitem{Dittrich:2000zu}
W.~Dittrich and H.~Gies,
%
Springer Tracts Mod.\ Phys.\  {\bf 166}, 1 (2000).


\bibitem{Tsai:1974fa}
W.~y.~Tsai and T.~Erber,
Phys.\ Rev.\ D {\bf 10}, 492 (1974).

\bibitem{Daugherty:1984tr}
J.~K.~Daugherty and A.~K.~Harding,
Astrophys.\ J.\  {\bf 273}, 761 (1983).

 
\bibitem{Cameron:1993mr}
  R.~Cameron {\it et al.} [BFRT Collaboration],
  Phys.\ Rev.\  D {\bf 47} (1993) 3707.

\bibitem{Zavattini:2007ee}
  E.~Zavattini {\it et al.}  [PVLAS Collaboration],
  arXiv:0706.3419 [hep-ex].

\bibitem{Chen:2006cd}
  S.~J.~Chen, H.~H.~Mei and W.~T.~Ni,
  Mod.\ Phys.\ Lett.\  A {\bf 22}, 2815 (2007)
  [arXiv:hep-ex/0611050].

\bibitem{Robilliard:2007bq}
  C.~Robilliard {\em et al.}, 
  Phys.\ Rev.\ Lett.\ {\bf 99}, 190403 (2007); 
  F.~Bielsa {\it et al.},
  arXiv:0911.4567 [physics.optics].


\bibitem{Pugnat:2006ba}
  P.~Pugnat {\it et al.}, CERN-SPSC-2006-035, CERN-SPSC-P-331, (2006); 
%
  P.~Pugnat {\it et al.}  [OSQAR Collaboration],
  Phys.\ Rev.\  D {\bf 78}, 092003 (2008)
  [arXiv:0712.3362 [hep-ex]].



\bibitem{Maiani:1986md}
  L.~Maiani, R.~Petronzio and E.~Zavattini,
  Phys.\ Lett.\ B {\bf 175}, 359 (1986); 
%
  G.~Raffelt and L.~Stodolsky,
  Phys.\ Rev.\ D {\bf 37}, 1237 (1988).


\bibitem{Gies:2006ca}
  H.~Gies, J.~Jaeckel and A.~Ringwald,
  Phys.\ Rev.\ Lett.\  {\bf 97}, 140402 (2006)
  [arXiv:hep-ph/0607118]; 
%
\bibitem{Ahlers:2006iz}
  M.~Ahlers, H.~Gies, J.~Jaeckel and A.~Ringwald,
  Phys.\ Rev.\  D {\bf 75}, 035011 (2007).

\bibitem{Ahlers:2007rd}
  M.~Ahlers, H.~Gies, J.~Jaeckel, J.~Redondo and A.~Ringwald,
  Phys.\ Rev.\  D {\bf 76}, 115005 (2007);
  Phys.\ Rev.\  D {\bf 77}, 095001 (2008)
  [arXiv:0711.4991 [hep-ph]].



\bibitem{Heinzl:2006xc}
  T.~Heinzl {\em et al.}, 
  Opt.\ Commun.\  {\bf 267}, 318 (2006)
  [arXiv:hep-ph/0601076].

\bibitem{Marx:2011ab}
B.~Marx, {\em et al.},
Opt.\ Commun. {\bf  284}, 915  (2011).

\bibitem{DiPiazza:2005jc}
  A.~Di Piazza, K.~Z.~Hatsagortsyan and C.~H.~Keitel,
  Phys.\ Rev.\  D {\bf 72}, 085005 (2005); 
  A.~Di Piazza, K.~Z.~Hatsagortsyan and C.~H.~Keitel,
  Phys.\ Rev.\ Lett.\  {\bf 97}, 083603 (2006)
  [arXiv:hep-ph/0602039].
  A.~Di Piazza, A.~I.~Milstein and C.~H.~Keitel,
  arXiv:0704.0695 [hep-ph]; .
  C.~Muller, A.~Di Piazza, A.~Shahbaz, T.~Burvenich, J.~Evers, K.~Hatsagortsyan and C.~Keitel,
  Laser Phys.\  {\bf 18}, 175 (2008).


\bibitem{Schutzhold:2008pz}
  R.~Schutzhold, H.~Gies and G.~Dunne,
  Phys.\ Rev.\ Lett.\  {\bf 101}, 130404 (2008)
  [arXiv:0807.0754 [hep-th]]; 
%
  G.~V.~Dunne, H.~Gies and R.~Schutzhold,
  Phys.\ Rev.\  D {\bf 80}, 111301 (2009)
  [arXiv:0908.0948 [hep-ph]];
%
%
  M.~Ruf, G.~R.~Mocken, C.~Muller, K.~Z.~Hatsagortsyan and C.~H.~Keitel,
  Phys.\ Rev.\ Lett.\  {\bf 102}, 080402 (2009)
  [arXiv:0810.4047 [physics.atom-ph]];
%
  T.~Heinzl, A.~Ilderton and M.~Marklund,
  arXiv:1002.4018 [hep-ph].
%
\bibitem{DiPiazza:2009py}
  A.~Di Piazza, E.~Lotstedt, A.~I.~Milstein and C.~H.~Keitel,
  Phys.\ Rev.\ Lett.\  {\bf 103}, 170403 (2009)
  [arXiv:0906.0726 [hep-ph]];
%
  S.~S.~Bulanov, V.~D.~Mur, N.~B.~Narozhny, J.~Nees and V.~S.~Popov,
  Phys.\ Rev.\ Lett.\  {\bf 104}, 220404 (2010)
  [arXiv:1003.2623 [hep-ph]];
  M.~Orthaber, F.~Hebenstreit and R.~Alkofer,
  arXiv:1102.2182 [hep-ph].


\bibitem{Hebenstreit:2009km}
  F.~Hebenstreit, R.~Alkofer, G.~V.~Dunne and H.~Gies,
  Phys.\ Rev.\ Lett.\  {\bf 102}, 150404 (2009)
  [arXiv:0901.2631 [hep-ph]]; 
  arXiv:0910.4457 [hep-ph].
\bibitem{Mackenroth:2010jk}
  F.~Mackenroth, A.~Di Piazza and C.~H.~Keitel,
  Phys.\ Rev.\ Lett.\  {\bf 105}, 063903 (2010)
  [arXiv:1001.3614 [physics.acc-ph]]; 
  C.~K.~Dumlu and G.~V.~Dunne,
   ``The Stokes Phenomenon and Schwinger Vacuum Pair Production in
  Phys.\ Rev.\ Lett.\  {\bf 104}, 250402 (2010)
  [arXiv:1004.2509 [hep-th]].


\bibitem{Tommasini:2009nh}
  D.~Tommasini, A.~Ferrando, H.~Michinel and M.~Seco,
  JHEP {\bf 0911}, 043 (2009)
  [arXiv:0909.4663 [hep-ph]].

\bibitem{King:10ab}
B.~King, A.~Di Piazza, and C.H.~Keitel, Nature Photon. {\bf 4}, 92 (2010).


\bibitem{Blaschke:2011af}
  D.~B.~Blaschke, G.~Roepke, V.~V.~Dmitriev, S.~A.~Smolyansky and A.~V.~Tarakanov,
  arXiv:1101.6021 [physics.plasm-ph]; 
  G.~Gregori {\it et al.},
  High Energy Dens.\ Phys.\  {\bf 6}, 166 (2010)
  [arXiv:1005.3280 [hep-ph]].

\bibitem{Harvey:2009ry}
  C.~Harvey, T.~Heinzl, A.~Ilderton,
  Phys.\ Rev.\  {\bf A79}, 063407 (2009).
  [arXiv:0903.4151 [hep-ph]]; 
%
  T.~Heinzl, D.~Seipt, B.~Kampfer,
  Phys.\ Rev.\  {\bf A81}, 022125 (2010).
  [arXiv:0911.1622 [hep-ph]].


\bibitem{Marklund:2008gj}
  M.~Marklund and J.~Lundin,
  Eur.\ Phys.\ J.\  D {\bf 55}, 319 (2009)
  [arXiv:0812.3087 [hep-th]].

\bibitem{Dunne:2008kc}
  G.~V.~Dunne,
  Eur.\ Phys.\ J.\  D {\bf 55}, 327 (2009)
  [arXiv:0812.3163 [hep-th]].

\bibitem{Heinzl:2008an}
  T.~Heinzl, A.~Ilderton,
  Eur.\ Phys.\ J.\  {\bf D55}, 359-364 (2009).
  [arXiv:0811.1960 [hep-ph]].


\bibitem{Gies:2008wv}
  H.~Gies,
  Eur.\ Phys.\ J.\  D {\bf 55}, 311 (2009)
  [arXiv:0812.0668 [hep-ph]].

\bibitem{Homma:2010jc}
  K.~Homma, D.~Habs and T.~Tajima,
  arXiv:1006.4533 [quant-ph].

\bibitem{Dobrich:2010hi}
  B.~Dobrich and H.~Gies,
  JHEP {\bf 1010}, 022 (2010)
  [arXiv:1006.5579 [hep-ph]]; 
  B.~Dobrich and H.~Gies,
  arXiv:1010.6161 [hep-ph].

\bibitem{Heinzl:2009zd}
  T.~Heinzl, A.~Ilderton, M.~Marklund,
  Phys.\ Rev.\  {\bf D81}, 051902 (2010).
  [arXiv:0909.0656 [hep-ph]].


\bibitem{Batalin:1971au}
  I.~A.~Batalin and A.~E.~Shabad,
  Zh.\ Eksp.\ Teor.\ Fiz.\  {\bf 60}, 894 (1971).

\bibitem{Ritus:1972ky}
  V.~I.~Ritus,
   ``Radiative corrections in quantum electrodynamics with intense field and
  Annals Phys.\  {\bf 69}, 555 (1972).


\bibitem{Cover:1974ij}
  R.~A.~Cover and G.~Kalman,
  Phys.\ Rev.\ Lett.\  {\bf 33}, 1113 (1974).


\bibitem{Urrutia:1977xb}
  L.~F.~Urrutia,
  Phys.\ Rev.\  D {\bf 17}, 1977 (1978).

\bibitem{Artimovich:1990qb}
  G.~K.~Artimovich,
   ``Properties of the photon polarization operator in an electric field:
  Sov.\ Phys.\ JETP {\bf 70}, 787 (1990)
  [Zh.\ Eksp.\ Teor.\ Fiz.\  {\bf 97}, 1393 (1990)].



\bibitem{Schubert:2000yt}
  C.~Schubert,
  Nucl.\ Phys.\ B {\bf 585}, 407 (2000)
  [arXiv:hep-ph/0001288].

\bibitem{Dittrich:2000wz}
  W.~Dittrich and R.~Shaisultanov,
  Phys.\ Rev.\  D {\bf 62}, 045024 (2000)
  [arXiv:hep-th/0001171].

\bibitem{Konar:2001mp}
  S.~Konar,
  Int.\ J.\ Mod.\ Phys.\  A {\bf 17}, 1055 (2002)
  [arXiv:hep-ph/0111104].


\bibitem{VillalbaChavez:2010bp}
  S.~Villalba-Chavez,
  Phys.\ Lett.\  B {\bf 692}, 317 (2010)
  [arXiv:1008.0547 [hep-th]].

\bibitem{Nikishov:1970br}
  A.~I.~Nikishov,
  Nucl.\ Phys.\  B {\bf 21}, 346 (1970).

\bibitem{Cangemi:1995ee}
  D.~Cangemi, E.~D'Hoker and G.~V.~Dunne,
  Phys.\ Rev.\  D {\bf 52}, 3163 (1995)
  [arXiv:hep-th/9506085].


\bibitem{Kim:2000un}
  S.~P.~Kim and D.~N.~Page,
  Phys.\ Rev.\  D {\bf 65}, 105002 (2002)
  [arXiv:hep-th/0005078].

\bibitem{Halpern:1977he}
  M.~B.~Halpern, A.~Jevicki and P.~Senjanovic,
   ``Field Theories In Terms Of Particle-String Variables: Spin, Internal
  Phys.\ Rev.\  D {\bf 16}, 2476 (1977); 
%
  Z.~Bern and D.~A.~Kosower,
  Nucl.\ Phys.\  B {\bf 379}, 451 (1992).
%
\bibitem{Strassler:1992zr}
  M.~J.~Strassler,
  Nucl.\ Phys.\  B {\bf 385}, 145 (1992)
  [arXiv:hep-ph/9205205]; 
%
\bibitem{Schmidt:1993rk}
  M.~G.~Schmidt and C.~Schubert,
  Phys.\ Lett.\  B {\bf 318}, 438 (1993)
  [arXiv:hep-th/9309055].

\bibitem{Schubert:2001he}
  C.~Schubert,
  Phys.\ Rept.\  {\bf 355}, 73 (2001)
  [arXiv:hep-th/0101036].

\bibitem{Gies:2001zp}
  H.~Gies and K.~Langfeld,
  Nucl.\ Phys.\  B {\bf 613}, 353 (2001)
  [arXiv:hep-ph/0102185]; 
%
  Int.\ J.\ Mod.\ Phys.\  A {\bf 17}, 966 (2002)
  [arXiv:hep-ph/0112198].

\bibitem{Langfeld:2002vy}
  K.~Langfeld, L.~Moyaerts and H.~Gies,
  Nucl.\ Phys.\  B {\bf 646}, 158 (2002)
  [arXiv:hep-th/0205304].

\bibitem{Gies:2005bz}
  H.~Gies and K.~Klingmuller,
  Phys.\ Rev.\  D {\bf 72}, 065001 (2005)
  [arXiv:hep-ph/0505099].

\bibitem{Gies:2003cv}
  H.~Gies, K.~Langfeld and L.~Moyaerts,
  JHEP {\bf 0306}, 018 (2003)
  [arXiv:hep-th/0303264]; 
%
  H.~Gies and K.~Klingmuller,
  Phys.\ Rev.\ Lett.\  {\bf 97}, 220405 (2006)
  [arXiv:quant-ph/0606235]; 
  A.~Weber and H.~Gies,
  Phys.\ Rev.\ Lett.\  {\bf 105}, 040403 (2010)
  [arXiv:1003.0430 [hep-th]].

\bibitem{Dunne:2009zz}
  G.~Dunne, H.~Gies, K.~Klingmuller and K.~Langfeld,
  JHEP {\bf 0908}, 010 (2009)
  [arXiv:0903.4421 [hep-th]].

\bibitem{Kharzeev:2006wg}
  D.~Kharzeev and K.~Tuchin,
  Phys.\ Rev.\  A {\bf 75}, 043807 (2007)
  [arXiv:hep-ph/0611133].

\bibitem{Gies:2005sb}
  H.~Gies, J.~Sanchez-Guillen and R.~A.~Vazquez,
  JHEP {\bf 0508}, 067 (2005)
  [arXiv:hep-th/0505275].

\end{thebibliography}
\end{document}